\documentclass[onecolumn,amsmath,amssymb,12pt,superscriptaddress,nofootinbib]{revtex4}
\usepackage[bookmarks,linktocpage, colorlinks=true, plainpages = false, citecolor = blue,  linkcolor=blue, urlcolor = blue,
filecolor = blue]{hyperref}

\usepackage{graphicx}
\usepackage{subfigure}
\usepackage{dcolumn}
\usepackage{bm}
\usepackage{braket}
\usepackage{verbatim} 
\usepackage[]{inputenc}
\usepackage{cancel}
\usepackage[usenames,dvipsnames]{color}
\usepackage{xfrac}

\input epsf
\def\gsim{ \lower .75ex \hbox{$\sim$} \llap{\raise .27ex \hbox{$>$}} }
\def\lsim{ \lower .75ex \hbox{$\sim$} \llap{\raise .27ex \hbox{$<$}} }

\newcommand{\be}{\begin{equation}}
\newcommand{\ee}{\end{equation}}
\newcommand{\ba}{\begin{aligned}}
\newcommand{\ea}{\end{aligned}}
\newcommand{\bea}{\begin{eqnarray}}
\newcommand{\eea}{\end{eqnarray}}
\renewcommand{\d}{\mathrm{d}}
\def\al{\alpha}
\def\b{\beta}

\def\Gb{\bar{\Gamma}}

\def\de{\delta}
\def\z{\zeta}

\def\cb{\bar{\chi}}
\def\s{\sigma}
\def\sb{\bar{\sigma}}
\def\sd{\dot{\sigma}}

\def\sp{\sigma^{\perp}}
\def\ep{\epsilon}
\def\f{\phi}
\def\fb{\bar{\phi}}

\def\O{\Omega}

\def\th{\theta}
\def\tb{\bar{\theta}}

\def\r{\rho}
\def\rb{\bar{\rho}}

\def\z{\zeta}

\def\p{\partial}
\def\ebsi{\bar{e}_{\s}}
\def\ebs{\bar{e}_{s}}
\def\eb{\bar{e}}
\def\td{\dot\theta}
\def\Rb{\bar{R}}
\def\Vb{\bar{V}}

\newcommand{\na}{\nabla}
\newcommand{\cL}{\mathcal{L}}

\newcommand{\cD}{\mathcal{D}}
\newcommand{\h}{\frac{1}{2}}

\newcommand{\parenthnewln}{\right.\\&\left.\quad\quad{}} 
\newcommand{\parenthnewlnln}{\right.\right.\\&\left.\left.\quad\quad{}} 
\newcommand{\parenthnewlnbigg}{\bigg.\\&\bigg.\quad\quad{}} 
\newcommand{\parenthnewlnlnbigg}{\bigg.\right.\\&\left.\bigg.\quad\quad{}}

\begin{document}
\allowdisplaybreaks
\begin{titlepage}

\title{The Non-Minimal Ekpyrotic Trispectrum}

\author{Angelika Fertig}
\email[]{angelika.fertig@aei.mpg.de}
\author{Jean-Luc Lehners}
\email[]{jlehners@aei.mpg.de}

\affiliation{Max--Planck--Institute for Gravitational Physics (Albert--Einstein--Institute), 14476 Potsdam, Germany}

\begin{abstract}

\vspace{.3in}
\noindent 
Employing the covariant formalism, we derive the evolution equations for two scalar fields with non-canonical field space metric up to third order in perturbation theory. These equations can be used to derive predictions for local bi- and trispectra of multi-field cosmological models. Our main application is to ekpyrotic models in which the primordial curvature perturbations are generated via the non-minimal entropic mechanism. In these models, nearly scale-invariant entropy perturbations are generated first due to a non-minimal kinetic coupling between two scalar fields, and subsequently these perturbations are converted into curvature perturbations. Remarkably, the entropy perturbations have vanishing bi- and trispectra during the ekpyrotic phase. However, as we show, the conversion process to curvature perturbations induces local non-Gaussianity parameters $f_{NL}$ and $g_{NL}$ at levels that should be detectable by near-future observations. In fact, in order to obtain a large enough amplitude and small enough bispectrum of the curvature perturbations, as seen in current measurements, the conversion process must be very efficient. Interestingly, for such efficient conversions the trispectrum parameter $g_{NL}$ remains negative and typically of a magnitude ${\cal O}(10^2) - {\cal O}(10^3),$ resulting in a distinguishing feature of non-minimally coupled ekpyrotic models.
\end{abstract}
\maketitle

\end{titlepage}
\tableofcontents

\clearpage
\section{Introduction}

In order to further our understanding of the beginning of the universe we can learn from at least two main sources; electromagnetic radiation and gravitational waves.
So far, all our knowledge is derived from the first, presenting us with a picture of the density distribution in the universe after photons from the surface of last scattering have been emitted. On the one hand, satellites like PLANCK \cite{Ade:2015xua, Ade:2015lrj, Ade:2015ava} have probed the Cosmic Microwave Background (CMB) radiation to exceedingly great detail. Surveys of the large-scale structure of the universe are quickly catching up, to the point where they might be able to rival the precision of the CMB maps in the near future, as our understanding of structure formation is continuously improving \cite{Alvarez:2014vva}. Further information will come from the detection (or absence) of primordial gravitational waves. In this light, it remains as important as ever to understand the predictions of cosmological models of the early universe.

The most popular theory of the early universe is inflation -- see \cite{Baumann:2009ds} for a review. Simple models of inflation lead to rather clear predictions: the fact that the inflaton must roll down a very flat potential implies that it is approximately a free field. This in turn implies a spectrum of density perturbations that is Gaussian to high accuracy, implying that both the bispectrum and trispectrum are expected to be very small. More complicated models can however be designed, involving multiple fields and/or higher derivative kinetic terms, such that essentially all potential combinations of observations can be matched. One may hope that this uncomfortable fact may be circumvented if additional constraints on model building, arising from the combination with particle physics or eventually quantum gravity, will become available. In the meantime, it is interesting to observe that simple inflationary models also typically predict primordial gravitational waves at observable levels, so that the current non-observation already starts to rule out a number of long-favoured models \cite{Ade:2015lrj}. 

A complicating feature of (most) inflationary models is the process of eternal inflation, by which the universe is turned into a multiverse of infinitely many ``pocket universes'' with different physical properties. In this way, even a single model of inflation can lead to an infinite number of possible outcomes. Despite many attempts, the process of eternal inflation and the resulting non-predictivity of inflation remain serious open problems of early universe cosmology, and they motivate the investigation of alternative models \cite{Ijjas:2013vea}. 

In the present paper we will be interested in ekpyrotic models, which form an alternative to inflation in that they can solve the flatness and horizon problems of the early universe, while also generating primordial density perturbations -- see \cite{Lehners:2008vx} for a review. Moreover, they are not plagued by the runaway behaviour of eternal inflation \cite{Johnson:2011aa,Lehners:2012wz}. However, they also present one big challenge: the ekpyrotic phase is a contracting phase assumed to precede the big bang and in order to obtain a complete model one must be able to explain a bounce linking the contracting phase to the currently expanding phase. Such bounces are not fully understood yet, but this is an active field of research in which significant progress is being made -- see for instance \cite{Creminelli:2007aq,Buchbinder:2007ad,Easson:2011zy,Cai:2012va,Koehn:2013upa,Gielen:2015uaa}. A distinguishing feature of ekpyrotic models is that they do not amplify gravitational waves \cite{Boyle:2003km} (except for a small effect at second order in perturbation theory, where the density perturbations act as a source \cite{Baumann:2007zm}). For this reason, we are all the more motivated to try to understand the predictions for the primordial density perturbations in great detail. In the present paper, we will therefore calculate the non-Gaussian corrections to the primordial density fluctuations, and in particular we will extend our previous treatment of the bispectrum/3-point function \cite{Fertig:2013kwa} to the trispectrum/4-point function.

The precise model we are interested in is the non-minimally coupled entropic mechanism, first proposed by Qiu, Gao and Saridakis \cite{Qiu:2013eoa}, and by Li \cite{Li:2013hga}, and generalised in \cite{Ijjas:2014fja}. Here, nearly scale-invariant entropy perturbations are generated by a field-dependent coupling between two scalar field kinetic terms. Subsequently, these entropy perturbations can then be converted into curvature perturbations. This model was shown in \cite{Fertig:2013kwa} to lead to a vanishing bispectrum during the ekpyrotic phase. Our aim in the present paper is twofold: we would like to understand the predictions of this model for the trispectrum, and investigate the effect of the conversion mechanism on both the bispectrum and trispectrum in detail. Note that, even though the scalar field space is now endowed with a non-trivial metric, the model does not contain higher derivative kinetic terms. For this reason, only the non-Gaussianities of local form are relevant, and these can be calculated from the classical evolution on large scales. Thus, we will first have to develop cosmological perturbation theory up to third order (since we are interested in the trispectum), for the case of a non-trivial field space metric and for models with two scalar fields. This will extend the existing treatment up to second order in perturbation theory of Renaux-Petel and Tasinato \cite{RenauxPetel:2008gi}, as well as the existing development of third order perturbation theory for trivial field space metrics \cite{Lehners:2009ja}. Note that this part of our paper is entirely general, and may be used for applications to any two-field cosmological models with arbitrary field space metrics (i.e. to general two-field non-linear sigma models).
 
What we find is that the non-minimally coupled ekpyrotic phase also leads to a precisely vanishing trispectrum, but that the conversion process has a crucial impact on the final predictions for the bispectrum and trispectrum of the curvature perturbations. In particular, we find that the conversion process must be very efficient in order for these models to be in agreement with current limits on the bispectrum parameter $f_{NL}.$ Interestingly, such efficient conversions then lead to a non-trivial prediction for the trispectrum non-linearity parameter $g_{NL},$ which is expected to be negative and of a magnitude of several hundred typically. This is thus an observational signature to look out for in future observations of the cosmic microwave background.

\section{Covariant formalism and perturbation theory up to second order} \label{section:covform}

We will start by reviewing the covariant formalism for cosmological perturbation theory, up to second order in perturbations and for a theory comprising two scalar fields (with non-trivial field space metric). Readers familiar with these results may proceed to the next section, where new results at third order will be presented. 

In the present work, we will adopt the $1+3$ covariant formalism developed by Langlois and Vernizzi \cite{Langlois:2005ii, Langlois:2005qp, Langlois:2006iq, Langlois:2006vv}, which was inspired by earlier works of Ellis and Bruni \cite{Ellis:1989jt,Bruni:1991kb} and Hawking \cite{Hawking:1966qi}. This formalism builds on the insight that in a purely time-dependent background metric (in particular in Friedmann-Lemaitre-Robertson-Walker (FLRW) spacetimes) spatial derivatives of scalar quantities are automatically gauge-invariant. The formalism allows one to derive rather compact all-orders evolution equations for cosmological perturbations, which, with suitable care, may then be expanded up to the desired order in perturbation theory.  

We study the cosmological fluctuations of a system of two non-minimally coupled scalar fields, i.e. two scalar fields with a non-trivial field space metric (but minimally coupled to gravity). The action of such systems is of the form
\be \label{eq:cL}
S = \int \mathrm{d}^4 x \sqrt{-g} \left( \frac{1}{2} R   - \h G_{IJ}(\phi^K) \nabla_{a} \phi^I \nabla^{a} \phi^J  - V(\phi^K) \right)\,,
\ee
where the indices $I,J,K=1,2$ label the two scalar fields (in our later examples we will also write $\phi^1=\phi, \phi^2=\chi$). The field space metric and its inverse can be used to lower and raise field space indices, respectively, e.g. ${\f}_I = G_{IJ} {\f}^J$.
Such actions were studied by Renaux-Petel and Tasinato \cite{RenauxPetel:2008gi} up to second order in perturbation theory and, for trivial field space metric $G_{IJ} = \delta_{IJ}$, the formalism was extended to third order by Renaux-Petel and one of us \cite{Lehners:2009ja}. Considering theories with two scalars fields is conceptually of importance, as two-scalar theories admit both adiabatic/curvature and entropic/isocurvature perturbations. The extension to having more than two fields is then straightforward, as the presence of additional fields simply augments the number of independent entropic perturbations.

\subsection{Covariant formalism}
Let us consider a spacetime, with metric $g_{ab}$, where a congruence of cosmological observers is defined by an a priori arbitrary unit timelike vector $u^a = \d x^a/ \d \tau \; (\textrm{with} \,\, u_a u^a = -1)$, where $\tau$ denotes the proper time. The spatial projection tensor orthogonal to the four-velocity $u^a$ is then given by
\be \label{def:hab}
h_{ab}\equiv g_{ab} + u_a u_b, \quad \quad
(h^{a}_{\ b} h^b_{\ c} = h^a_{\ c}, \quad h_a^{\ b} u_b=0). 
\ee
To describe the time evolution, we make use of the Lie derivative w.r.t. $u^a$, i.e. the covariant definition of the time derivative. It is defined for a generic covector $Y_a$ by (see e.g. \cite{Wald:1984r})
\be \label{def:Lie}
\dot Y_a\equiv \cL_u Y_a \equiv u^b \nabla_b Y_a + Y_b \nabla_a u^b,
\ee
and will be denoted by a \textit{dot}, as has been customary in works on the covariant formalism. For scalar quantities, the covariant time derivative reduces to
\be 
\dot f = u^b \nabla_b f. 
\ee 
To describe perturbations in the covariant approach, we project the covariant derivative orthogonally to the four-velocity $u^a$; this spatial projection of the covariant derivative will be denoted by $D_a$. For a generic tensor, its definition is
\be \label{def:Datensor}
D_a T_{b \dots}^{\ c \dots} \equiv h_a^{\ d}h_b^{\ e}\dots h^{\ c}_f \dots \nabla_d T_{e \dots}^{\ f \dots}. 
\ee 
Again, for the case of a scalar, this simplifies,
\be \label{def_Dascalar} 
D_a f \equiv h_a^{\ b} \nabla_b f = \partial_a f + u_a \dot f. 
\ee 

The covariant derivative of any time-like unit vector field $u^a$ can be decomposed
uniquely as follows
\be \label{decomposition} 
\nabla_b u_a=\sigma_{ab}+\omega_{ab}+{1\over 3}\Theta h_{ab}-a_a u_b,
\ee 
with the (trace-free and symmetric) shear tensor
$\sigma_{ab}$ and the (antisymmetric) vorticity tensor
$\omega_{ab}$. The volume expansion, $\Theta$, is defined by
\be  \label{def:Theta}
\Theta \equiv \nabla_a u^a, 
\ee 
where the integrated volume expansion, $\al$, along $u^a$,
\be
\al \equiv \frac{1}{3} \int \d \tau \, \Theta \quad\;\; (\Theta = 3 \dot{\al}),
\ee
can be interpreted as the number of e-folds of evolution of the scale factor measured along the world-line of a cosmological observer with four-velocity $u^a$ since $\Theta/3$ corresponds to the local Hubble parameter. 
The acceleration vector is given by
\be \label{def:accvector}
 a^a \equiv u^b\nabla_b u^a .
\ee  

Finally, it is always possible to decompose the total energy-momentum tensor as
\be \label{Tab2} 
T_{ab} =(\rho + p) u_a u_b + q_a u_b + u_a q_b + g_{ab} p + \pi_{ab}, 
\ee 
where
$\rho$, $p$, $q_a$ and $\pi_{ab}$ are the energy density, pressure,  momentum and anisotropic stress tensor, respectively, as measured in the frame defined by $u^a$.

\subsection{Two scalar fields with non-trivial field space metric}

The energy momentum tensor derived from the action (\ref{eq:cL}) is then
\be \label{Tab}
T_{a b} = G_{IJ} \nabla_{a} \phi^I \nabla_{b} \phi^J  + g_{a b} (- \h G_{IJ} \nabla_{c} \phi^I \nabla^{c} \phi^J  - V). 
\ee
Comparing to the decomposition in (\ref{Tab2}) one finds for the energy density, pressure, momentum and anisotropic stress, respectively,
\bea \label{eq:rho}
\rho &\equiv& T_{ab} u^a u^b =T_{00} u^0 u^0 = \dot{\phi}^I \dot{\phi}_I + \h G_{IJ} \nabla_{c} \phi^I \nabla^{c} \phi^J  + V , \label{rho}\\
p &\equiv& \frac{1}{3} h^{ac} T_{ab} h^b_{\, c} = \frac{1}{3} G_{IJ} D_a \phi^I D^a \phi^J - \h G_{IJ} \nabla_{c} \phi^I \nabla^{c} \phi^J  - V, \label{p}\\
q_a &\equiv& -u^b T_{bc} h^c_{\, a}=  - \dot{\phi}_I D_a \phi^I \approx - u^0 T_{0i}, \label{qa}\\
\pi_{ab} &\equiv&   h_a^{\, c} T_{cd} h^d_{\, b} - p \, h_{ab} = G_{IJ} D_a \phi^I D_b \phi^J -\frac{h_{ab}}{3}  G_{IJ} D_c \phi^I D^c \phi^J.
\label{piab}
\eea
The equation of motion for the scalar fields is obtained by varying the action w.r.t. the fields themselves,
\be \label{eq:eomphiI}
\ba
0 =& G_{IJ} \nabla_a \nabla^a \phi^J + \Gamma_{IJK} \nabla_a \phi^K \nabla^a \phi^J - V_{,I} \\
=& \ddot{\phi}^I+ \Gamma^I_{JK} \left(\dot{\phi}^J \dot{\phi}^K - D_a \phi^J D^a \phi^K \right) + \Theta \dot{\phi}^I + G^{IJ} V_{,J} - D_a D^a \phi^I - a^b D_b \phi^I,
\ea
\ee
where $\Gamma_{IJK}=G_{IL} \Gamma^L_{JK}=\h \left( G_{IJ,K} + G_{IK,J} - G_{JK,I} \right)$ and the second equality above makes use of equations (\ref{def:hab}, \ref{def:Datensor}-\ref{def:accvector}).

We introduce the following derivatives of field space vectors in curved coordinates in order to simplify notation. The spacetime derivative, given by
\be
\cD_a A^I \equiv \nabla_a A^I + \Gamma^I_{JK} \nabla_a \phi^J A^K,
\ee
is used to define a time derivative in field space,
\be \label{eq:cDu}
\cD_u A^I \equiv u^a \cD_a A^I = \dot{A}^I + \Gamma^I_{JK} \dot{\phi}^J A^K,
\ee
and a spatially projected derivative in field space,
\be
\cD_{\perp a} T_{b \dots}^{I \; c \dots} \equiv h_a^d h_b^e \dots h_f^c \dots \cD_d T_{e \dots}^{I \; f \dots}.
\ee
We can then rewrite the evolution equation (\ref{eq:eomphiI}) in a more concise form as
\be \label{eq:eomphiI2}
\cD_u \dot{\phi}^I + \Theta \dot{\phi}^I + G^{IJ} V_{,J} - \cD_{\perp a} \left( D^a \phi^I \right) - a^a D_a \phi^I =0.
\ee

\smallskip

In the two-field case it is convenient to introduce a particular basis in field space which consists of an adiabatic and an entropic unit vector. This decomposition was first introduced in \cite{Gordon:2000hv} for two fields in the linear theory. The generalisation to multiple fields is discussed in \cite{Langlois:2008mn, GrootNibbelink:2000vx} for the linear case and in \cite{Rigopoulos:2005xx} for the nonlinear theory. The adiabatic unit vector, denoted by $e_{\s}^I$, is defined in the direction of the velocity of the two fields, i.e. \textit{tangent} to the field space trajectory. The entropic unit vector, denoted by $e_s^I$, is defined along the direction \textit{orthogonal} to it (w.r.t. $G_{IJ}$), namely
\be \label{eq:basis}
e_{\s}^I \equiv  \frac{\dot{\phi}^I}{\dot{\s}},
\qquad 
G_{IJ} e_s^I e_s^J =1, 
\qquad 
G_{IJ} e_s^I e_{\s}^J =0,
\ee
with 
\be
\sd \equiv \sqrt{G_{IJ} \dot{\phi}^I \dot{\phi}^J}.
\ee
Note that this is only a short-hand notation, i.e. $\sd$ is generally {\it not} the derivative along $u^a$ of a scalar field $\sigma$. Furthermore, we introduce the quantity $\td$ to express the time evolution of the basis vectors,
\be \label{eq:cDue}
\cD_u e_{\s}^I \equiv \dot{\theta} e_{s}^I, \qquad \cD_u e_{s}^I \equiv - \dot{\theta} e_{\s}^I,
\ee
where $\cD_u e_{\al}^I = \dot{e}_{\al}^I + \Gamma^I_{JK} \dot{\s} e_{\s}^J e_{\al}^K, (\al=\s,s)$ is given by the definition in (\ref{eq:cDu}). Again, $\dot{\theta}$ is {\it not} the derivative along $u^a$ of an angle $\th$, although such an angle can be defined for a trivial field space metric \cite{Langlois:2006vv}.

\smallskip

Making use of the basis (\ref{eq:basis}), we can then introduce two linear combinations of the scalar field gradients and thus define two covectors by analogy with the similar
definitions in the linear context \cite{Langlois:2008mn}: the {\it adiabatic} and {\it
entropic} covectors, denoted by $\s_a$ and $s_a$, respectively, and given by
\bea
\s_a    \equiv  e_{\sigma I} \nabla_a \phi^I
\label{def:sigmaa}, \\
 s_a \equiv e_{s I} \nabla_a \phi^I
  \label{def:sa}.
\eea
By definition, the entropic covector $s_a$ is orthogonal to the four-velocity $u^a$, i.e. $u^a s_a = 0$. By contrast, the adiabatic covector $\s_a$ contains a longitudinal component: $u^a \s_a = \dot{\s}$. At any location in spacetime, one may think of $\sigma_i$ as describing perturbations in the total energy density (and thus perturbations in the expansion/contraction history of the universe), and of $s_i$ as describing perturbations in the relative contributions of the two scalar fields to the total energy density. 

\smallskip

A covariant generalisation of the comoving energy density perturbation is given by the covector
\be \label{def:ep}
\ep_a \equiv D_a \rho - \frac{\dot{\rho}}{\dot{\s}}\sp_a,  
\ee 
where $\sp_a \equiv e_{\s I} D_a \f^I = \s_a + \sd u_a$ is the spatially projected version of (\ref{def:sigmaa}). It has been shown in \cite{Langlois:2006vv} that if the shear is negligible on large scales, so is $\ep_a \approx 0$.

Then, in our two-field system, the full evolution equation of the entropy covector $s_a$ can be expressed on large scales (i.e. to leading order in spatial gradients) as \cite{RenauxPetel:2008gi}
\be \label{eq:eomsa}
\ddot{s}_a + \Theta \dot{s}_a +\left( V_{;ss} + 3 \dot{\th}^2 + \dot{\sigma}^2  e_{s}^{I} e_{s}^{J} e_{\s}^{K} e_{\s}^{L} R_{IKJL} \right) s_a \approx - 2 \frac{\dot{\th}}{\dot{\s}} \ep_a,
\ee
with
\be
V_{,s} = e_{s}^{I} V_{,I}, \;\;\;\;\;\; V_{;ss} = e_{s}^{I} e_{s}^{J} \cD_I \cD_J V\;\;\;\;\;\; \text{and} \;\;\;\;\;\;  
\cD_I \cD_J V \equiv V_{,IJ} - \Gamma^K_{IJ} V_{,K},
\ee
and where $R^I_{\ KLJ} = \p_J \Gamma^I_{KL} - \p_L \Gamma^I_{KJ} + \Gamma^I_{JP} \Gamma^P_{KL} - \Gamma^I_{LP} \Gamma^P_{KJ}$ is the Riemann tensor associated with the metric $G_{IJ}$.
An equality valid only on large scales will be denoted by $\approx$.

\smallskip

It is a well-known result that in cosmological models with a single scalar field the curvature perturbation is conserved on large scales \cite{Wands:2000dp}. However, when a second field is present, entropic perturbations may arise and these can source the curvature perturbation on large scales \cite{Gordon:2000hv}. We will now derive a particularly simple and useful form of the evolution equation for the comoving curvature perturbation on large scales, extending known versions \cite{Lyth:2004gb,Buchbinder:2007at,Lehners:2009qu} to the case of having a non-trivial field space metric. We will work in comoving gauge and take the background to be described by a flat FLRW metric. On large scales the perturbed metric can be written as
\be
ds^2 = - dt^2 + a(t)^2 e^{2 \z(t,x^i)} dx^i dx_i, 
\ee
where $\z$ denotes the comoving curvature perturbation and can be thought of as a local perturbation in the scale factor. We will denote derivatives w.r.t. physical time $t$ with primes. In our model \eqref{eq:cL} the equation of continuity is given by
\be
\rho' + 3 (H +\z') (\rho + p) =0,
\ee
with the background (denoted by overbars) satisfying
\be
\bar{\rho}' + 3 \bar{H} (\bar{\rho} + \bar{p}) =0.
\ee
On co-moving hypersurfaces the energy density is uniform, $\rho=\bar{\rho}$ (and hence also $H=\bar{H}$ because of the Friedmann equation). We then obtain
\be \label{eq:curvpertp}
\z' = - \bar{H} \frac{\de p}{\bar{\rho} + \bar{p} + \de p},
\ee
where $\de p \equiv p(t,x^i)-\bar{p}(t)$. Since, by definition, $\de \rho = 0$ on these hypersurfaces, we can immediately relate the pressure perturbation to a perturbation in the potential, $\de p = - 2 \de V|_{\de \rho = 0}.$
Plugging this relation into equation (\ref{eq:curvpertp}), we obtain a compact expression for the evolution of the comoving curvature perturbation on large scales, writing $\bar{H}=H$,
\be \label{eq:zeta}
\zeta' = \frac{2 H \de V}{\bar{\s}'^2 - 2 \de V},
\ee
which is valid to all orders in perturbation theory.

\smallskip

In the following subsection, we will introduce a coordinate system. The evolution equations for the entropy perturbation (\ref{eq:eomsa}) and the comoving curvature perturbation (\ref{eq:zeta}) can then be straightforwardly translated into the linearized and second-order perturbation equations, while we derive new results at third order in the following section. For convenience, we have collected various background as well as first- and second-order expressions that will be used in the rest of this paper in appendix \ref{section:appendixB}.

\subsection{Perturbation theory up to second order}

We introduce coordinates $x^{\mu}=(t,x^i)$ to describe an almost-FLRW spacetime, in order to relate the covariant formalism to the more familiar coordinate based approach. 
We will denote a partial derivative with respect to the cosmic time t by a prime, i.e. $' = \p/\p t$, since the dot is already reserved for the Lie derivative (\ref{def:Lie}). 
Fields are expanded without factorial factors:
\be
X(t,x^i) = \bar{X}(t)+\de X^{(1)}(t,x^i)+\de X^{(2)}(t,x^i)+\de X^{(3)}(t,x^i).
\ee
Quantities with an over-bar like $\bar{X}$ are evaluated on the background, first order quantities like $\de X^{(1)}$ solve the linearised equations of motion, second order quantities like $\de X^{(2)}$ the quadratic equations, and so on. In the following, we drop the superscript $^{(1)}$ for perturbations at linear order when the meaning is unambiguous.
For simplicity we choose $u^{a}$ such that $u_i = 0$. In appendix \ref{app:metric} we show how $u_0$ is then determined in terms of metric quantities.

\smallskip

We start by presenting the definitions of the adiabatic and entropic perturbations up to second order. By expanding Eqs.~(\ref{def:sigmaa}) and (\ref{def:sa}) up to second order, one finds,
for $\sigma_i$ and $s_i$ respectively,
\bea \label{eq:desigma}
\de \s_i=\partial_i \de \s, \qquad \de \s \equiv \bar{e}_{\s I} \delta \phi^I \\
\label{eq:des}
\de s_i=\partial_i \de s, \qquad \de s \equiv \bar{e}_{s I} \delta \phi^I.
\eea
at linear order and
\bea
\label{eq:desigma2i} 
\de \s_i^{(2)} &\equiv& \p_i \de \s^{(2)} + \frac{\tb'}{\sb'} \de \s \p_i \de s -\frac{1}{\sb'} V_i, \label{si_i} \\
\label{eq:des2i}  
\delta s_i^{(2)}  &\equiv& \p_i \de s^{(2)} + \frac{\de \s}{\sb'}  \p_i \de s',
\eea
at second order \cite{RenauxPetel:2008gi}, with
\bea
\label{eq:desigma2} 
\de \s^{(2)} &\equiv&  \bar{e}_{\s I}  {\de \phi^{I (2)}}  
+ \h \eb_{\s I} \Gb^I_{KL} \bar{e}_{\al}^K  \bar{e}_{\b}^L \de \s^{\al} \de \s^{\b}
+ \frac{1}{2 \sb'} \de s \de s'   \\
\label{eq:des2} 
\de s^{(2)} &\equiv& \bar{e}_{s I}  {\de \phi^{I (2)}}  
+ \h \eb_{s I} \Gb^I_{KL} \bar{e}_{\al}^K  \bar{e}_{\b}^L \de \s^{\al} \de \s^{\b}
-\frac{\de \s}{\sb'}  \left( \de s' +
\frac{\tb'}{2} \de \s \right),
\eea
where the inverse zweibeine are defined via $\de \phi^I = e_{\al}^{\ I} \de \s^{\al}$ and $\al= (\s, s)$.
The curved nature of the field space metric manifests itself in the appearance of the terms with Christoffel symbols in $\de \s^{(2)}$ and $\de s^{(2)}$.
It is convenient to introduce the spatial vector
\be\label{def:Vi}
V_i \equiv \h (\de s \p_i \de s' - \de s' \p_i \de s), 
\ee
which vanishes when $\de s$ and $\de s'$ have the same spatial dependence. Since relative spatial gradients are heavily suppressed for super-Hubble modes both in inflationary and in ekpyrotic models, $\de s'$ and $\de s$ indeed obtain the same spatial dependence, i.e. $\de s'=g(t) \de s,$ to high precision.

\smallskip

The gauge transformation of a tensor ${\bf T}$ generated by a vector field $\xi^a$ is given by the exponential map \cite{Malik:2008im} 
\be
 {\bf T} \rightarrow e^{\cL_{\xi}} {\bf T}\,.
\ee 
With the perturbative expansion $\xi=\sum_n \frac{1}{n!} \, \xi_{(n)}$, the first and second-order perturbations of a tensor  ${\bf T}$ are then found to transform as
\cite{Bruni:1996im} 
\be \label{gauge_transformation}
{\bf \delta T}^{(1)}\rightarrow {\bf \de T}^{(1)} +\cL_{\xi_{(1)}} {\bf T}^{(0)}, \quad \;\;
{\bf \de T}^{(2)} \rightarrow {\bf \de T}^{(2)}  + \cL_{\xi_{(1)}} {\bf \de T}^{(1)} + \left( \cL_{\xi_{(2)}} + \h \cL_{\xi_{(1)}}^2 \right) {\bf T}^{(0)}. 
\ee 
Using these relations, it can easily be verified that the entropic perturbations $\de s^{(1),(2)}$ are gauge-invariant. 
The adiabatic perturbations, however, are not gauge-invariant, but they have been defined such that setting them to zero is equivalent to going to comoving gauge, on large scales. This can be seen by expanding the momentum density $q_i$ given by (\ref{qa}), which ought to vanish in comoving gauge:\be \label{deqi} 
\de q_i =-\p_i \left(\sb' \de \s \right) 
\ee 
at linear order, and
\bea \label{deqi2} 
\de q_i^{(2)}  =-\p_i \left[\sb' \de \s^{(2)} +\h \frac{\sb''}{\sb'} \de \s^2 +
\tb' \de \s \de s \right] -\frac{1}{\sb'} \de \ep \p_i \de \s + V_i, 
\eea 
at second order. As already mentioned, $V_i \approx0$ on large scales for the models we are interested in, and therefore setting the adiabatic perturbations to zero (as a
gauge choice) corresponds to adopting
comoving gauge on super-Hubble scales.

\smallskip

The equations of motion of the scalar fields were presented as (\ref{eq:eomphiI}), which for the background can be rewritten as
\be \label{eq:bgeom}
\bar{G}_{IJ} \Box \fb^J + \Gb_{IKL} \p_{\mu} \fb^K \p^{\mu} \fb^L - \Vb_{,I} =0. 
\ee
Substituting $\fb^{J'}= \sb' \ebsi^J$  and using (\ref{eq:cDue}), they read 
\be
\eb_{\s I} \left( \sb'' + 3 H \sb' \right) + \eb_{s I} \sb' \tb' + \bar{V}_{,I} = 0.
\ee
Multiplying with $\ebsi^I$ and $\ebs^I$, we obtain the background equations of motion for $\s$ and $s$, respectively:
\be
\sb'' + 3 H \sb' + V_{,\s} = 0,
\ee
and
\be
 \sb' \tb' + V_{,s} = 0.
\ee

\smallskip

Expanding the equation of motion for $s_a$ (\ref{eq:eomsa}) to linear order gives
\be \label{eq:eomdes}
\de s'' + 3H \de s' +  \left( \bar{V}_{;ss} + 3 \tb'^2 + \sb'^2 \bar{e}_{s}^{I} \bar{e}_{s}^{J} \bar{e}_{\s}^{K} \bar{e}_{\s}^{L}   \bar{R}_{IKJL} \right)  \de s = - \frac{2\tb'}{\sb'} \de \ep \approx 0,
\ee
where we have used
\be
\bar{\Theta} = 3 H.
\ee

At second order, we get
\be \label{eq:eomdes2}
\ba
&\de s^{(2)''} + 3H \de s^{(2)'} + \left( \bar{V}_{;ss} + 3 \tb'^2 + \sb'^2 \bar{e}_{s}^{I} \bar{e}_{s}^{J} \bar{e}_{\s}^{K} \bar{e}_{\s}^{L}   \bar{R}_{IKJL} \right)  \de s^{(2)} \approx - \frac{\tb'}{\sb'} \de s'^2 \\
&- \frac{2}{\sb'} \left[\tb'' + \frac{\Vb_{,\s} \tb'}{\sb'} - \frac{3}{2} H \tb'  \right] \de s \de s' 
+ \left[ -\h \Vb_{;sss} + \frac{5 \Vb_{;ss} \tb' }{\sb'} + \frac{9 \tb'^3}{\sb'} 
\parenthnewln 
+  \bar{e}_{s}^{I} \bar{e}_{s}^{J} \bar{e}_{\s}^{K} \bar{e}_{\s}^{L}  \left( \sb' \tb' \bar{R}_{IKJL} -\h \sb'^2 \ebs^N \cD_N \bar{R}_{IKJL}  \right) \right] \de s^2 -\frac{2 \tb'}{\sb'} \de \ep^{(2)},
\ea
\ee
where we have used $V_i \approx 0$ on large scales in the second term on the RHS,
and $\de \ep^{(2)} \approx 0$.
The equation for the entropy perturbation forms a closed system; on large scales, it evolves independently of the adiabatic component.

\smallskip

In comoving gauge, expanding the expression for the curvature perturbation (\ref{eq:zeta}) up to second order, we have
\be \label{eq:curvpertV}
\ba
\z' &= \frac{2 H  \de V}{\sb'^2 - 2 \de V } \, \stackrel{\de \s = 0}{\approx} \,  \frac{2 H}{\sb'^2} \left[ \de V^{(1)} + \de V^{(2)} + \frac{2}{\sb'^2} \left(\de V^{(1)} \right)^2  \right],
\ea
\ee
where the $\de \s = 0$ statement above the $\approx$ sign indicates that the equations are valid in comoving gauge. Using equations (\ref{eq:deV1})-(\ref{eq:deV2}), we obtain
\be \label{eq:zeta1comgauge}
\z^{(1)'} \stackrel{\de \s = 0}{\approx} - \frac{2 H \tb' \de s}{\sb' }
\ee
at first order, and
\be \label{eq:zeta2comgauge}
\z^{(2)'} \stackrel{\de \s = 0}{\approx}  \frac{2 H }{\sb'^2 } \left[ - \sb' \tb' \de s^{(2)}  - \frac{\Vb_{,\s}}{2 \sb'} \de s \de s' + \left( \h \Vb_{;ss} + 2 \tb'^2 \right) \de s^2
\right]
\ee
at second order.
It becomes clear that the curvature perturbation is sourced by the entropy perturbation.

In the next section, we will derive the corresponding third-order equations, which are needed for the study of the primordial trispectra of cosmological perturbations.

\section{Deriving the third order equations of motion} \label{sec:O3eom}

We are now in a position to present our main technical developments: we use the covariant formalism to derive the third-order evolution equations for the entropy and the curvature perturbations for two scalar fields with a non-trivial field space metric. These equations will then allow us to calculate and make predictions for the trispectrum of current ekpyrotic models.

The covariant formalism has the advantage of allowing one to derive simple all-orders evolution equations for the adiabatic and entropic co-vectors. However, given that the general all-orders definitions of the adiabatic and entropic convectors are rather implicit and formal, in using the covariant formalism to make actual predictions a non-trivial step consists in identifying the proper definitions of adiabatic and entropic fluctuations up to the desired order in perturbation theory. Once these definitions are at hand, it becomes a straightforward exercise to expand the all-orders equations up to the desired order. Thus our first and main task is to find the appropriate definitions of adiabatic and entropic perturbations at third order. Expanding Eq. (\ref{def:sa}) at third order using Eqs. (\ref{eq:deesJ}) and (\ref{eq:de2esJ}), one obtains
\be \label{eq:des3i}
\ba
\de s_i^{(3)} =\, &\p_i \de s^{(3)} + \frac{\de \s}{\sb'} \p_i \de s^{(2)'} 
+ \frac{\de \s^{(2)}}{\sb'} \p_i \de s' - \frac{\sb''}{2 \sb'^3} \de \s^2 \p_i \de s'  + \frac{\de \s^2}{2 \sb'^2} \p_i \de s'' \\
& + \frac{1}{2 \sb'^2} \left( \de s' + 2 \tb' \de \s\right) \de s \p_i \de s'  - \p_i \left( \frac{1}{6 \sb'^2} \de s \de s'^2 \right) \\ & + \frac{1}{6} \ebs^I \ebs^J \ebsi^K \ebsi^L  \Rb_{IKJL} \left(  \de \s^2 \p_i \de s - \de \s \de s \p_i \de \s \right) 
\\ & + \frac{1}{3} \eb_{sI}\left[ - \p_M \Gb^I_{KL} + \Gb^I_{KP} \Gb^P_{LM} + \Gb^I_{LP} \Gb^P_{KM} \right] \ebs^L \left( \ebsi^M \ebs^K - \ebsi^K \ebs^M \right) \p_i \left( \de \s \de s^2 \right) 
\ea
\ee
where we have defined
\be \label{eq:des3}
\ba
\de s^{(3)} \equiv\, &\eb_{s I} \de \f^{(3) I} - \frac{\de \s^{(2)}}{\sb'} \left( \de s' + \tb' \de \s \right) - \frac{\de \s}{ \sb'}  \de s^{(2)'} - \frac{\de \s^2}{2 \sb'^2} \left( \tb'^2 \de s + \de s'' - \frac{\sb''}{\sb'} \de s' \right) \\ & - \frac{\de \s^3}{6 \sb'} \left( \frac{\tb'}{\sb'} \right)' - \frac{\tb'}{2 \sb'^2} \de \s \de s \de s' + \frac{1}{6 \sb'^2} \de s \de s'^2
\\ &+ \eb_{s I} \Gb^I_{KL} \left( \ebsi^L \de \s + \ebs^L \de s \right)\left[\ebsi^K \left( \de \s^{(2)} - \frac{1}{2 \sb'} \de s \de s' \right) 
+ \ebs^K \left( \de s^{(2)} + \frac{\de \s}{\sb'} \left(\de s' + \frac{\tb'}{2} \de \s\right)\right)\right]
\\ & - \frac{1}{6} \left( - \p_M \Gb^I_{KL} + \Gb^I_{KP} \Gb^P_{LM} + \Gb^I_{LP} \Gb^P_{KM} \right) \eb_{sI} \times \\
&\times \left[ \left( \ebsi^M \de \s + \ebs^M \de s \right) \left( \ebsi^L \de \s + \ebs^L \de s \right) \left( \ebsi^K \de \s + \ebs^K \de s \right)  
 \parenthnewln +\de \s \de s \left( \ebsi^L \de \s + 2 \ebs^L \de s \right) \left( \ebsi^M \ebs^K - \ebsi^K \ebs^M \right) \right].
\ea
\ee
Using the transformation of the third-order perturbations of a tensor  ${\bf T}$, given by \cite{Bruni:1996im} 
\be
{\bf \de T}^{(3)} \rightarrow {\bf \de T}^{(3)} + \cL_{\xi_{(1)}} {\bf \de T}^{(2)} + \left( \cL_{\xi_{(2)}} + \h \cL_{\xi_{(1)}}^2 \right) {\bf \de T}^{(1)} + \left( \cL_{\xi_{(3)}} + \cL_{\xi_{(1)}} \cL_{\xi_{(2)}}+ \frac{1}{6} \cL_{\xi_{(1)}}^3 \right) {\bf T}^{(0)},
\ee
one can show that the entropy perturbation as defined in (\ref{eq:des3}) is gauge-invariant. Note that, compared to the earlier work \cite{Lehners:2009ja}, we have added the gauge-invariant term $\frac{1}{6 \sb'^2} \de s \de s'^2$ to the definition of $\de s^{(3)}$ (and correspondingly subtracted off its derivative from $\de s_i^{(3)}$). This improved definition is motivated by our considerations of ekpyrotic models in section \ref{sec:ekpexamples}, as we will further discuss there. Moreover, in appendix \ref{sec:O3T0} we will present additional arguments that the term that we are adding to the definition of the entropic perturbation is the only sensible one\footnote{This new definition does not change the results of \cite{Lehners:2009ja}, as our new definition differs from the old one by a gauge-invariant term.}. Apart from this small modification, the present definition now also includes terms due to the curvature of field space.

On large scales, our new definition leads to an extremely simple relationship between the covector $\de s_i^{(3)}$ and the entropic perturbation $\de s^{(3)}$: in comoving gauge we have
\be \label{eq:des3icomgauge}
\ba
\de s_i^{(3)} &\stackrel{\de \s = 0}{\approx} \p_i \de s^{(3)} + \frac{1}{2 \sb'^2} \de s \de s' \p_i \de s'  - \frac{1}{6 \sb'^2} \p_i \left(\de s \de s'^2  \right)
\\ & \;\,=
\p_i \de s^{(3)} + \frac{1}{3 \sb'^2}  \de s' V_i
\\ &\;\, \approx \p_i \de s^{(3)}
\ea
\ee
with
\be \label{eq:des3comgauge}
\ba
\de s^{(3)} \stackrel{\de \s = 0}{\equiv}\, &\eb_{s I} \de \f^{(3) I} + \frac{1}{6 \sb'^2} \de s \de s'^2
+ \eb_{s I} \Gb^I_{KL}  \ebs^L \de s \left[\ebs^K  \de s^{(2)} - \frac{1}{2 \sb'}  \ebsi^K \de s \de s'  
 \right]  \\
&+ \frac{1}{6} \eb_{sI} \ebs^J \ebs^K \ebs^L \left[  \p_J \Gb^I_{KL} - 2 \Gb^I_{JP} \Gb^P_{KL}  \right]  \de s^3,
\ea
\ee
where we have simplified the last term due to the symmetry in the vielbeine.

\smallskip

The adiabatic perturbation $\de \s$ is not a gauge-invariant variable, so there is more freedom in choosing a definition. Expanding Eq. (\ref{def:sigmaa}) using Eqs. (\ref{eq:deesigmaJ}) and (\ref{eq:de2esigmaJ}), we obtain
\be \label{eq:desi3i}
\ba
\de \s_i^{(3)} =\, &\p_i \de \s^{(3)} + \frac{\tb'}{\sb'} \left( \de \s \p_i \de s^{(2)} + \de \s^{(2)} \p_i \de s \right)
+ \frac{\de \s}{\sb'^2} \left( \de s' + \tb' \de \s \right) \p_i \de s' + \frac{1}{2 \sb'^2} \left( \de s' + \tb' \de \s \right)^2 \p_i \de \s \\
&+ \left[  \left( \frac{\tb'}{\sb'}  \right)' \frac{\de \s}{2} - \frac{1}{\sb'^2} \left( \Vb_{;ss} + 2 \tb'^2 \right) \de s + \frac{\Vb_{,\s}}{\sb'^2} \de s' \right] \de \s \p_i \de s -\frac{1}{\sb'} V_i^{(3)} - \frac{\tb'}{3 \sb'^2} \de s V_i \\
&+ \left( \ebsi^K \ebs^L + \ebs^K \ebsi^L \right) \ebsi^J \de \s \de s \p_i \left( \ebsi^I \de \s + \ebs^I \de s \right) \left[ \h \left(\bar{G}_{IP,J} - \bar{G}_{IJ,P}\right) \Gb^P_{KL} \parenthnewln
- \bar{G}_{JP,L} \Gb^P_{IK} + \frac{1}{4} \left( \bar{G}_{KL,IJ} - \bar{G}_{IK,LJ} - \bar{G}_{IL,KJ} + 2 \bar{G}_{IJ,KL} \right) \right],
\ea
\ee
with
\be\label{eq:desi3}
\ba
\de \s^{(3)} \equiv\, &\eb_{\s I} \de \f^{(3) I}+\frac{1}{2 \sb'} \left( \de s' \de s^{(2)} + \de s \de s^{(2)'} \right) + \frac{\tb'}{6 \sb'^2} \de s^2 \de s' \\
&+ \eb_{\s I} \Gb^I_{KL} \left( \ebsi^L \de \s + \ebs^L \de s \right)  
\left[\ebsi^K \left( \de \s^{(2)} - \frac{1}{2 \sb'} \de s \de s' \right) + \ebs^K \left( \de s^{(2)} + \frac{\de \s}{\sb'} \left(\de s' + \frac{\tb'}{2} \de \s\right)\right)\right]  \\
&+ \h  \ebs^I \ebs^J \ebsi^K \ebsi^L   \Rb_{IKJL} \de \s \de s^2 
+ \frac{1}{6} \eb_{\s I} \left( \ebsi^J \ebsi^K \ebsi^L \de \s^3 + \ebs^J \ebs^K \ebs^L \de s^3 \right) \left[ \p_J \Gb^I_{KL} -2 \Gb^I_{JP} \Gb^P_{KL} \right]. 
\ea
\ee
We have defined the natural generalisation of the third order non-local term $V_i$ as 
\be
V_i^{(3)} = \h \left( \de s^{(2)} \p_i \de s' + \de s \p_i  \de s^{(2)'} - \de s^{(2)'} \p_i \de s - \de s' \p_i \de s^{(2)} \right),
\ee
which again vanishes when the total entropy perturbation $\de s = \de s^{(1)} + \de s^{(2)}$ factorizes in terms of its time and spatial dependence, i.e. $\de s' = g(t) \de s$.
We can neglect it as such differences in spatial gradients are heavily suppressed on large scales in both inflationary and ekpyrotic models. 
For $\de \s = \de \s^{(2)} =0$ the adiabatic perturbation at third order reduces to
\be\label{eq:desi3comgauge}
\ba
\de \s^{(3)} \stackrel{\de \s = 0}{\approx}& \eb_{\s I} \de \f^{(3) I}+\frac{1}{2 \sb'} \left( \de s' \de s^{(2)} + \de s \de s^{(2)'} \right) + \frac{\tb'}{6 \sb'^2} \de s^2 \de s' \\
&+ \eb_{\s I} \Gb^I_{KL} \ebs^L \de s 
\left[\ebs^K  \de s^{(2)} - \frac{1}{2 \sb'} \ebsi^K \de s \de s'  \right]
+ \frac{1}{6} \eb_{\s I}  \ebs^J \ebs^K \ebs^L \de s^3 \left[ \p_J \Gb^I_{KL} -2 \Gb^I_{JP} \Gb^P_{KL} \right]. 
\ea
\ee
One may check that this is a useful definition of the adiabatic perturbation by expanding the momentum density (\ref{qa}) to third order and verifying that it vanishes on large scales, $\de q_i^{(3)} \stackrel{\de \s = 0}{\approx} 0$, in comoving gauge $\de \s = \de \s^{(2)} = \de \s^{(3)} =0.$

\smallskip

Now that we have the definitions of the adiabatic and entropic fluctuations, we can obtain their equations of motion. To this end, we expand the equation of motion for $s_a$ (\ref{eq:eomsa}) to third order, with the result
\be \label{eq:eomdes3_new}
\ba
&0 \approx \de s^{(3)''} + 3H \de s^{(3)'} + \left( \bar{V}_{;ss} + 3 \tb'^2 + \sb'^2 \bar{e}_{s}^{I} \bar{e}_{s}^{J} \bar{e}_{\s}^{K} \bar{e}_{\s}^{L}   \bar{R}_{IKJL} \right)  \de s^{(3)}  + 2 \frac{\tb'}{\sb'} \de s' \de s^{(2)'} \\
&+\left( \frac{2}{\sb'} \tb'' +\frac{2}{\sb'^2} \Vb_{,\s} \tb' - \frac{3}{\sb'} H \tb' \right) \left( \de s \de s^{(2)} \right)' \\
&+\left( \Vb_{;sss} - \frac{10}{\sb'} \Vb_{;ss} \tb'  - \frac{18}{\sb'} \tb'^3 +  \ebs^I \ebs^J \ebsi^K \ebsi^L  \left( -2 \sb' \tb' \Rb_{IKJL} + \sb'^2 \ebs^N \cD_N \Rb_{IKJL}  \right) \right) \de s \de s^{(2)} \\
&+ \frac{\Vb_{,\s}}{3 \sb'^3} \de s'^3 
+ \frac{1}{\sb'^2} \left[\frac{2}{3} \Vb_{;\s\s} + \frac{2 \Vb_{,\s}^2}{\sb'^2} + \frac{1}{\sb'} H \Vb_{,\s} -  \Vb_{;ss} - \frac{8}{3} \tb'^2 - \sb'^2 \bar{e}_{s}^{I} \bar{e}_{s}^{J} \bar{e}_{\s}^{K} \bar{e}_{\s}^{L}  \bar{R}_{IKJL}  \right] \de s \de s'^2 \\
&+ \left[ - \frac{22}{3\sb'^2} \tb' \tb'' - \frac{7}{6 \sb'} \Vb_{;ss\s} 
- \frac{11}{3 \sb'^3} \Vb_{;ss} \Vb_{,\s} - \frac{13}{3 \sb'^3} \Vb_{,\s} \tb'^2 - \frac{1}{\sb'^2} H \Vb_{;ss} +\frac{18}{\sb'^2} H \tb'^2 \parenthnewln 
- \frac{4 \Vb_{,\s} }{3 \sb'}  \bar{e}_{s}^{I} \bar{e}_{s}^{J} \bar{e}_{\s}^{K} \bar{e}_{\s}^{L}  \bar{R}_{IKJL} 
- \frac{\sb' }{6} \bar{e}_{s}^{I} \bar{e}_{s}^{J} \bar{e}_{\s}^{K} \bar{e}_{\s}^{L} \bar{e}_{\s}^{M} \cD_M  \bar{R}_{IKJL}   \right] \de s^2 \de s' \\
&+ \bigg[ \frac{1}{6} \Vb_{;ssss} - \frac{7}{3 \sb'} \Vb_{;sss} \tb' + \frac{5}{3 \sb'^2} \Vb_{;ss}^2 + \frac{19}{\sb'^2} \Vb_{;ss} \tb'^2 + \frac{24}{\sb'^2} \tb'^4 
\parenthnewlnbigg
+ \frac{1}{3} \ebs^I \ebs^J \ebsi^K \ebsi^L \left( \Rb_{IKJL}  \left( \Vb_{;ss} + \tb'^2 \right) 
-2 \sb' \tb'  \ebs^N \cD_N \Rb_{IKJL}  
\parenthnewlnlnbigg
+ \sb'^2 \ebs^N \ebs^Q \left(   \h \cD_Q \cD_N \bar{R}_{IKJL} -  \Rb_{IKJP} \Rb^P_{\; NLQ} + \Rb_{IKJL} \Rb^P_{\; NPQ} \right) 
\right)\bigg] \de s^3,
\ea
\ee
where we have used $V_i \approx 0$ and $V^{(3)}_i \approx 0$ on large scales. The equation of motion is fully covariant, as it should. Notice that upon the introduction of the extra term in the definition of $\de s^{(3)}$ in (\ref{eq:des3}) compared to \cite{Lehners:2009ja} the numerical factors of some of the terms have changed -- see appendix~\ref{sec:O3T0} for the equivalent equation of motion without the extra term. Moreover, the non-trivial field space metric manifests itself in the appearance of terms with Riemann tensors and their covariant derivatives. Just as was the case at lower orders, the large-scale equation for the entropy perturbation is closed at third order.

\smallskip

On large scales, the evolution of the curvature perturbation at third order is given by expanding (\ref{eq:zeta}) and using (\ref{eq:deV1})-(\ref{eq:deV3}), leading to
\be \label{eq:zeta3comgauge}
\ba
\z^{(3)'} &\stackrel{\de \s = 0}{\approx}   \frac{2 H}{\sb'^2} \left[ \de V^{(3)} + \frac{4}{\sb'^2} \de V^{(1)} \de V^{(2)} + \frac{4}{\sb'^4} \left(\de V^{(1)} \right)^3  \right] \\
&\;\,= \;\, \frac{2 H }{\sb'^2 } \left[ - \sb' \tb' \de s^{(3)}  - \frac{\Vb_{,\s}}{2 \sb'} \left( \de s \de s^{(2)} \right)'   + \left(  \Vb_{;ss} + 4 \tb'^2 \right) \de s \de s^{(2)}+ \frac{\tb'}{6 \sb'} \de s \de s'^2  \parenthnewln
+ \left( \frac{11}{6} \frac{\tb'\Vb_{,\s}}{\sb'^2}  - \frac{1}{2 \sb'} \Vb_{;s\s}  \right) \de s^2 \de s' + \left( \frac{1}{6} \Vb_{;sss} - 2 \frac{\tb' \Vb_{;ss}}{\sb'} -4 \frac{\tb'^3}{\sb'} \right) \de s^3
\right].
\ea
\ee
It is the third-order counterpart of equations (\ref{eq:zeta1comgauge}) and (\ref{eq:zeta2comgauge}) and shows how the adiabatic/curvature perturbations are sourced by entropic perturbations. As is apparent from the first line, once the potential $V$ becomes irrelevant, $\z$ is conserved on large scales. This is for instance the case in the approach to the bounce in ekpyrotic models.

\section{Ekpyrotic examples} \label{sec:ekpexamples}

\subsection{The non-minimally coupled ekpyrotic model}
The evolution equations derived in the previous section (\ref{sec:O3eom}) can be applied to any inflationary or ekpyrotic model described by two scalar fields with a non-trivial field space metric and a potential. In this paper we are chiefly interested in the ``non-minimal entropic mechanism'', which is a mechanism for generating ekpyrotic density perturbations. It was first proposed by Qiu, Gao and Saridakis \cite{Qiu:2013eoa} as well as by Li \cite{Li:2013hga}, and further developed and generalised in \cite{Fertig:2013kwa,Ijjas:2014fja}. The model contains two scalar fields: $\phi$ is assumed to have an ordinary kinetic term and a steep negative potential -- thus $\phi$ drives the ekpyrotic contracting phase. A second scalar, $\chi,$ is non-minimally coupled to $\phi$ such that in the ekpyrotic background it obtains nearly scale-invariant perturbations. Compared to the standard entropic mechanism, the model has the advantage that it does not require an unstable potential to generate nearly scale-invariant perturbations. In fact, in this model the potential need not depend on the second scalar $\chi$ at all during the ekpyrotic phase.
The entropic mechanism consists of a two-stage process: first nearly scale-invariant, Gaussian entropy perturbations are produced during the ekpyrotic phase, which are then converted into curvature perturbations in the subsequent kinetic phase by a bending in the field space trajectory. We will assume that the conversion process also occurs during the contracting phase of the universe. To complete the model, one may then consider both a prescription for initial conditions \cite{Battarra:2014xoa,Battarra:2014kga,Lehners:2015sia,Lehners:2015efa} and a non-singular bounce into the current expanding phase of the universe - see for instance \cite{Creminelli:2007aq,Buchbinder:2007ad,Cai:2012va,Koehn:2013upa} for a discussion of non-singular bounces and \cite{Xue:2013bva,Battarra:2014tga} for the proof that the perturbations generated during the contracting phase pass through such non-singular bounces unharmed. 

As just described, we will consider the case where the second scalar field $\chi$ is coupled to the first scalar $\f$ by a function $\O(\f)^2$, i.e. the field space metric and its inverse are given by
\be\label{eq:fieldspacemetric}
G_{IJ} = \begin{pmatrix}
  1  &  0 \\
  0  &  \O(\f)^2  \\
 \end{pmatrix},
\;\;\;\;\; \text{and} \;\;\;\;\;
G^{IJ} = \begin{pmatrix}
  1  &  0 \\
  0  &  \O(\f)^{-2}  \\
 \end{pmatrix}.
\ee
In a FLRW universe, the background equations of motion derived from the action (\ref{eq:cL}) are then
\bea
&& \fb'' + 3 H \fb' + \Vb_{,\f} - \O \O_{,\f} \cb'^2 = 0, \label{eq:phieom} \\
&& \cb'' + \left( 3 H + 2 \O^{-1} \O_{,\f} \fb' \right) \cb' + \O^{-2} \Vb_{, \chi} = 0,  \label{eq:chieom} \\
&& H^2 = \frac{1}{6} \left(\fb'^2 + \O^2 \cb'^2 + 2 \Vb \right).  \label{eq:fried1}
\eea

\smallskip
In the next subsection, in order to obtain the non-Gaussianity parameters we solve for the entropy and curvature perturbations, first analytically during the ekpyrotic phase, and then numerically for the conversion phase. Simplifications brought about by our choice of field space metric $G_{IJ}$ are detailed in appendix~\ref{sec:specificmodel}.

\subsection{The ekpyrotic phase}\label{sec:Ekpphase}

During the ekpyrotic phase, the potential is a function of $\phi$ alone, $V=V(\phi)$, and hence $V_{,\chi}=0$. From the background equation of motion for $\chi$ (\ref{eq:chieom}) it is immediately clear that $\chi'=0$ is a solution. 
The remaining background equations (\ref{eq:phieom}, \ref{eq:fried1}) then reduce to those for a single scalar in an ekpyrotic potential, $V(\f) = - V_0 e^{- \sqrt{2 \ep} \f}$, and they admit the scaling solution \cite{Khoury:2001wf}
\be
a \propto (-t)^{1/\ep}, \;\; \phi = 
\sqrt{\frac{2}{\ep}} \ln{\left[ - \left( \frac{V_0 \ep^2}{ (\ep -3)} \right)^{\h}  t\right]},
\ee
where $t$ is negative and runs from large negative towards small negative values. The fast-roll parameter $\ep \equiv  \fb'^2/(2H^2)$ is directly related to the equation of state $w=2\ep/3-1$ and for a successful ekpyrotic phase where the universe becomes flat and anisotropies are suppressed we need $\ep>3$. 
It has been shown in \cite{Ijjas:2014fja} that for any ekpyrotic equation of state it is possible to choose the potential and the kinetic coupling such that nearly scale-invariant entropy perturbations are produced.

\smallskip
We will now turn our attention to the perturbations, first reviewing the results at linear and second order. During the ekpyrotic phase, the curvature perturbations obtain a blue spectrum \cite{Lyth:2001pf} and moreover they are not amplified \cite{Tseng:2012qd,Battarra:2013cha}, such that we do not need to discuss them. More interesting are the entropic perturbations. In the constant $\chi$ background the entropic direction in field space is precisely the $\chi$ direction. 

The specification of comoving gauge, $\de \s^{(1)} =\de \s^{(2)} = 0$, translates directly to 
\be\label{eq:dephiekp}
\de \phi^{(1)}|_{\text{ekp}}=0
\ee
at linear order from (\ref{eq:desigma}), and
\be\label{eq:dephi2ekp}
\de \phi^{(2)}|_{\text{ekp}} = \frac{1}{2 \sb'} \left( \de s \de s' -  \O^{-1}  \O_{,\f} \bar{\f}'   \de s^2 \right)  = - \h \O^2 \bar{\f}'^{-1} \de \chi \de \chi'
\ee
at second order from (\ref{eq:desigma2}). With the definitions of the entropy perturbation at linear and quadratic order from equations (\ref{eq:des}) and (\ref{eq:des2}),
\be \label{eq:desek}
\de s|_{\text{ekp}} = - \O(\f) \de \chi,
\ee
and
\be \label{eq:des2ek}
\de s^{(2)}|_{\text{ekp}} = - \O(\f) \de \chi^{(2)},
\ee
the evolution equations for the entropy perturbation simplify significantly: at linear order, starting from (\ref{eq:eomdes}) we obtain
\be
\de s'' + 3 H \de s' + \left[ \O^{-1} \O_{,\f} \bar{V}_{,\f}  - \O^{-1} \O_{,\f\f} \bar{\f}'^2 \right] \de s \approx 0,
\ee
which rewritten in terms of $\de \chi$ and making use of the background equation for $\phi$ (\ref{eq:phieom}) becomes
\be \label{eq:eomdechiek}
\de \chi'' + \left( 3 H + 2 \O^{-1} \O_{,\f} \bar{\f}' \right) \de \chi'  \approx 0.
\ee
It is immediately clear that $\de \chi' = 0$ is a solution during the ekpyrotic phase\footnote{The solution at linear order is non-zero ($\de \chi^{(1)} = \text{constant}$) due to the quantization and associated amplification of the perturbations.}. This further simplifies our definitions; the second order perturbation in the first scalar field (\ref{eq:dephi2ekp}) vanishes,
\be
\de \phi^{(2)}|_{\text{ekp}} =0.
\ee

It is straightforward then to show that during the ekpyrotic phase, the equation of motion for the entropy perturbation at second order, given by (\ref{eq:eomdes2}), takes the same form as the first order one, namely
\be
\de s^{(2)''} + 3 H \de s^{(2)'} + \left[ \O^{-1} \O_{,\f} \bar{V}_{,\f}  - \O^{-1} \O_{,\f\f} \bar{\f}'^2 \right]  \de s^{(2)}  \approx 0.
\ee
There arises no source term for the second-order entropy perturbation $\de s^{(2)}$, and we have the trivial solution
\be
\de s^{(2)}|_{\text{ekp}} = 0,
\ee
generating no intrinsic non-Gaussianity for the entropy perturbations. By contrast, the entropy perturbations develop significant local non-Gaussian corrections in the standard entropic mechanism already during the ekpyrotic phase, due to the $\chi$-dependence of the potential \cite{Buchbinder:2007at,Koyama:2007if,Lehners:2007wc,Lehners:2008my,Lehners:2009ja,Lehners:2009qu,Lehners:2010fy}.

Having solved for the entropy perturbation, we can use equations (\ref{eq:zeta1comgauge}) and (\ref{eq:zeta2comgauge}) to obtain the evolution equation for the curvature perturbation at linear and quadratic order, respectively, as \cite{Fertig:2013kwa}
\be \label{eq:zeta1comek}
\zeta^{(1)'}|_{\text{ekp}} \approx 0,
\ee
noting that  $\tb'|_{\text{ekp}} =0$, and
\be \label{eq:zeta2comek}
\zeta^{(2)'}|_{\text{ekp}} \approx \frac{ H \bar{V}_{,\phi}}{\bar{\s}'^2} \left[ \bar{\s}'^{-1} \de s \de s' + \O^{-1} \O_{,\f}   \de s^2 \right] = -\frac{ H \bar{V}_{,\f}}{\bar{\f}'^3}  \O^2  \de \chi \de \chi' =0,
\ee
where the last equality follows from (\ref{eq:eomdechiek}).
Thus, during the ekpyrotic phase, no second-order curvature perturbation is generated, $f_{NL \,\text{integrated}} =0$ \cite{Fertig:2013kwa}. This becomes clear once one realises that the linearised solution, given by $\de \chi = \text{constant}$, behaves analogously to the background.

\smallskip

We can now apply our new results to extend this discussion to third order. During the ekpyrotic phase and with the field space metric given in (\ref{eq:fieldspacemetric}) the equation of motion~(\ref{eq:eomdes3_new}) at third order simplifies to
\be
\ba \label{eq:eomdes3ekp_new}
\de s^{(3)''} &+ 3 H \de s^{(3)'} + \left[ \O^{-1} \O_{,\f} \bar{V}_{,\f}  - \O^{-1} \O_{,\f\f} \bar{\f}'^2 \right]  \de s^{(3)}  \approx 0
\ea
\ee
allowing the solution
\be
\de s^{(3)}|_{\text{ekp}} = - \O \de \chi^{(3)} = 0.
\ee
Like at second order, no intrinsic non-Gaussianity for the entropy perturbations is generated at third order for this class of models. Note that if we had not added the gauge-invariant term $\frac{1}{6 \sb'^2} \de s \de s'^2$ to the definition of the third order entropy perturbation in (\ref{eq:des3}), then $\de s^{(3)}$ would have been non-zero. This would not have changed any results for physically measurable quantities, but it is clear that our present definition of the third order entropy perturbation is preferable to the older definition of \cite{Lehners:2009ja}, both on physical and aesthetic grounds.

The curvature perturbation at third order can be calculated by noting that during the ekpyrotic phase, $\Vb_{,\chi} = \tb' =0$, and hence
\be
\de V|_{\text{ekp}} = \Vb_{,\f} \de \f + \Vb_{,\f} \de \f^{(2)} + \h \Vb_{,\f\f} \de \f^2 + \Vb_{,\f} \de \f^{(3)} + \Vb_{,\f\f} \de \f \de \f^{(2)} + \frac{1}{6} \Vb_{,\f\f\f} \de \f^3 + \mathcal{O}(4),
\ee
which simplifies to
\be \label{eq:deVekcomgauge}
\de V|_{\text{ekp}} \stackrel{\de \s = 0}{\approx}  \Vb_{,\f} \de \f^{(3)} + \mathcal{O}(4),
\ee
in comoving gauge. From equation (\ref{eq:desi3comgauge}), we have that during the ekpyrotic phase (in comoving gauge on large scales) 
\be \label{eq:comgauge3}
\de \s^{(3)}|_{\text{ekp}} \stackrel{\de \s = 0}{\approx} - \de \f^{(3)} =0.
\ee
Thus there is no source for the curvature perturbation during the ekpyrotic phase,
\be
\z^{(3)'}|_{\text{ekp}} = \frac{2 H}{\sb'^2} \de V^{(3)} \stackrel{\de \s = 0}{\approx} \frac{2 H}{\sb'^2}  \Vb_{,\f} \de \f^{(3)} =0,
\ee
and at third order also the comoving curvature perturbation remains zero during the ekpyrotic phase, i.e. we have $g_{NL \,\text{integrated}} =0$.

In summary, we find that the ekpyrotic phase produces no local non-Gaussianity at all -- at least up to third order in perturbation theory -- both for the entropy and the curvature fluctuations. As we will now see, the conversion process of entropy into curvature fluctuations will change this result appreciably.

\subsection{The conversion phase}\label{sec:numericalconversion}
After the ekpyrotic phase has come to an end, during the subsequent kinetic phase the conversion from entropy to curvature perturbations is achieved by a bending in the field space trajectory. This bending occurs naturally in the heterotic M-theory embedding of the ekpyrotic/cyclic model \cite{Khoury:2001wf,Lehners:2006pu,Lehners:2006ir,Lehners:2007nb}, though other origins of such a bending may of course also be envisaged. The bending of the scalar field space trajectory can be modelled by having a repulsive potential (given a specific realisation of the cyclic model in heterotic M-theory, this repulsive potential can in principle be calculated \cite{Lehners:2007nb}). Here, in order to be general, we consider four different representative forms for the repulsive potential, namely 
\be
\ba \label{eq:Vrep}
V_{1,2}= \,v  \left[ x^{-2} + r \,x^{-6} \right]\!, \;\; v  \left[ \left( \sinh{x }\right)^{-2} + r \left( \sinh{x }\right)^{-4} \right]\!,
\ea
\ee
with $r=0, 1$ and where the dependence of the potential on $x=- \frac{\f}{2} + \frac{\sqrt{3} \, \chi}{2}$ expresses the fact that the repulsive potential forms an angle (here chosen to be $\pi/6$) with respect to the background trajectory.

\smallskip
One of the important parameters is the duration of the conversion process. We measure it by the number of e-folds $N$ of the evolution of $|aH|$ during conversion. That is, one e-fold of conversion corresponds to $a'(t_{\text{conv-end}}) =  e \cdot a'(t_{\text{conv-beg}})$. In our numerical studies we determine $N$ by determining the number of e-folds during which 90 percent of the total bending takes place, i.e. we require $\int_{t_{\text{conv-beg}}}^{t_{\text{conv-end}}} \tb' \d t / \int_{t_{\text{kin-beg}}}^{t_{\text{kin-end}}} \tb' \d t = 0.9$.
Conversions lasting about one e-fold correspond to what we call smooth conversions, while shorter conversions are sharper. We find that the results depend very significantly on the smoothness of conversion. 

As previously argued in \cite{Fertig:2013kwa}, in the non-minimal entropic mechanism the local bispectrum produced during the conversion process is small when the conversion is efficient (which corresponds to the conversion being smooth \cite{Lehners:2008my}). However, it is rather non-trivial to obtain such an efficient conversion process. This becomes clear when we analyse the equation of motion for $\chi$ given in (\ref{eq:chieom}), where the potential is now the repulsive potential modelling the conversion. Even small changes along the background trajectory (along $\s \sim \f$) lead to an enormous factor $\O^{-2} \sim e^{\f}$ multiplying the now non-zero $\chi$-derivative of the potential. This causes the background trajectory to be sharply deflected leading to an extremely inefficient conversion. So whenever the scalar curvature, given by
\be \label{eq:R}
R = -2 \frac{\O_{,\f\f}}{\O},
\ee
is significant, the conversion is highly inefficient. This has the consequence of leading to a small amplitude for the curvature perturbations, and large non-Gaussianities in clear contradiction with observations. What this means is that the field space metric, taken to be $\O = e^{-b\phi/2}$ during the ekpyrotic phase, has to become flatter again during the conversion process. Thus, in the same way as the potential turns off after the end of the ekpyrotic phase, the field space metric must progressively return to being trivial.

\subsubsection{Linearly decaying field space curvature}

\begin{figure}[h!]
\centering
\subfigure[\label{fig:Bessel_phi_chi} Field space trajectory]
  {\includegraphics[scale=0.55]{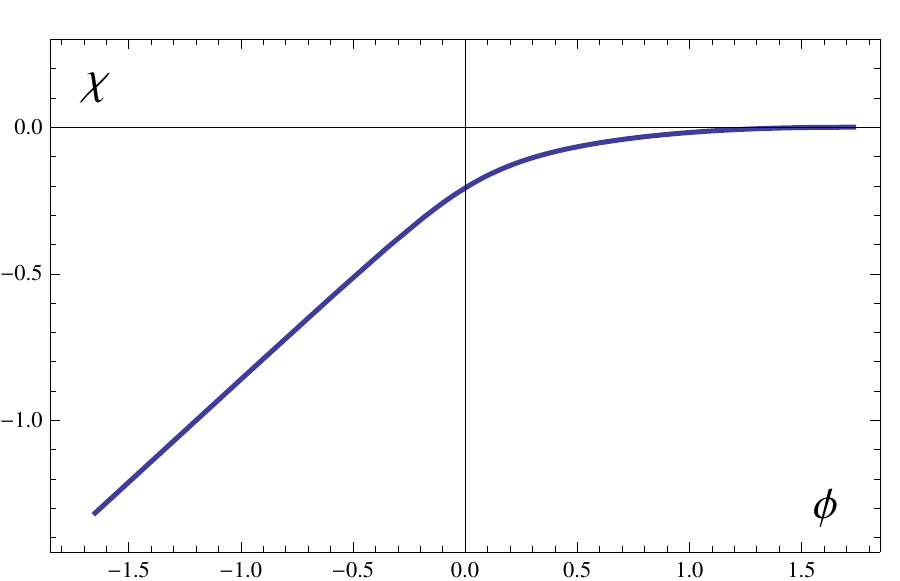}}
\subfigure[\label{fig:Bessel_Omega_phi} Field space metric]
  {\includegraphics[scale=0.55]{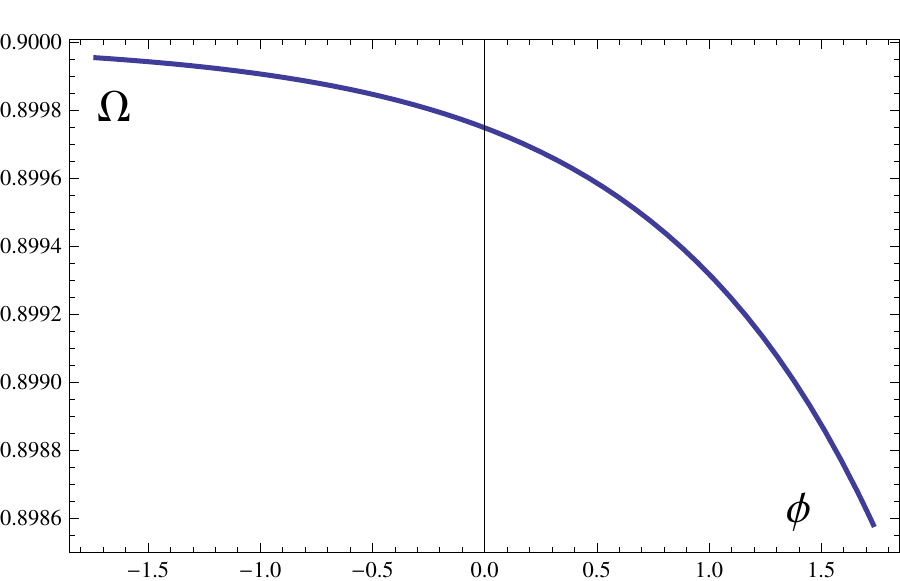}}
\subfigure[\label{fig:Bessel_R_t} Field space curvature]
  {\includegraphics[scale=0.534]{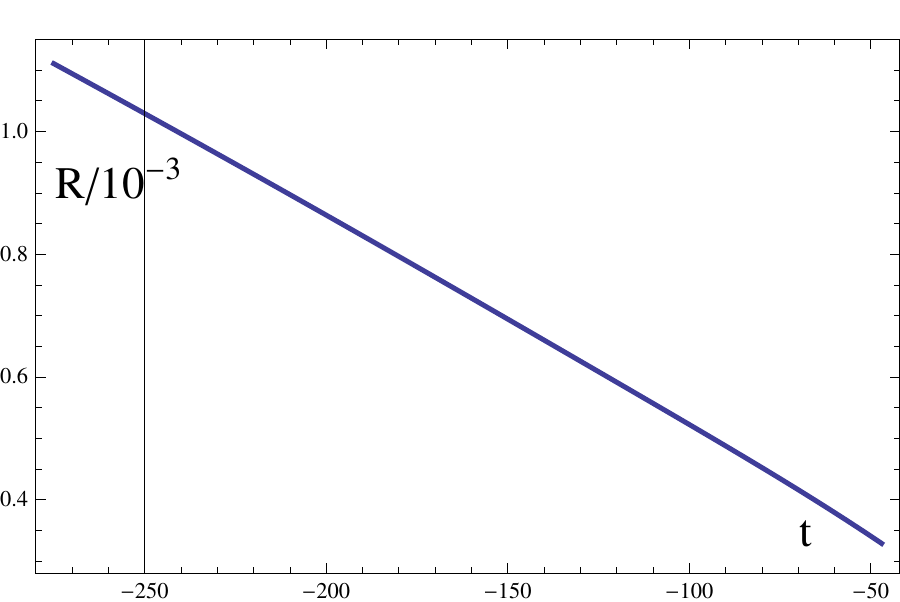}}
\caption{\label{fig:Bessel_Omega_phi_chi} The evolution of the fields, the field space metric ($\O = 1 - b \cdot I_0 (d \cdot \mathrm{e}^{c \f / 2})$) and the scalar (Ricci) curvature during one e-fold of conversion (from $t=-275$ to $t=-47$), plotted for the specified case $r=1$, $c=1$, $b=d=\frac{1}{10}$ giving $f_{\text{NL}}=-1.3$ and $g_{\text{NL}}=-544$.}
\end{figure}

Motivated by the previous discussion, we want to analyse cases where the field space metric returns to being trivial during the conversion process, after the end of the ekpyrotic phase. We will first concentrate on the case where the Ricci scalar of the field space decays linearly with time. This can be modelled by a kinetic coupling function of the form
\be \label{eq:OmegaBessel}
\O = 1 - b \cdot I_0 (d \cdot \mathrm{e}^{c \f / 2}),
\ee
where $I_0$ is a modified Bessel function of the first kind. This has the nice property that the scalar curvature (\ref{eq:R}) has a constant slope $R'$ and decays linearly, as seen in figure \ref{fig:Bessel_R_t}.  We have plotted $\O$ in figure \ref{fig:Bessel_Omega_phi} in the region of the conversion, and figure \ref{fig:Bessel_phi_chi} shows the bending of the trajectory for a typical smooth conversion lasting one e-fold. 

\begin{figure}[h!]
\centering
\subfigure[\ $f_{NL}$ vs. $R'$.]
  {\includegraphics[scale=0.85]{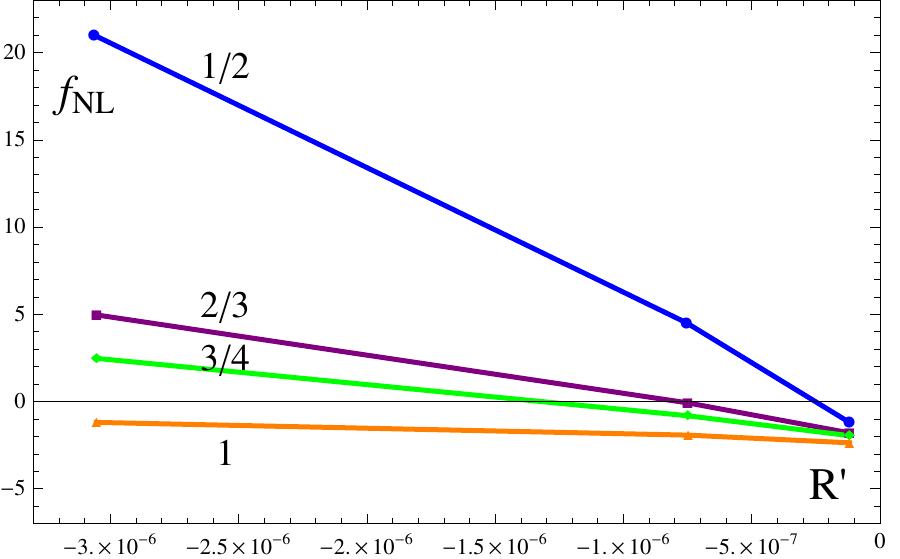}}
\subfigure[\ $g_{NL}$ vs. $R'$.]
  {\includegraphics[scale=0.85]{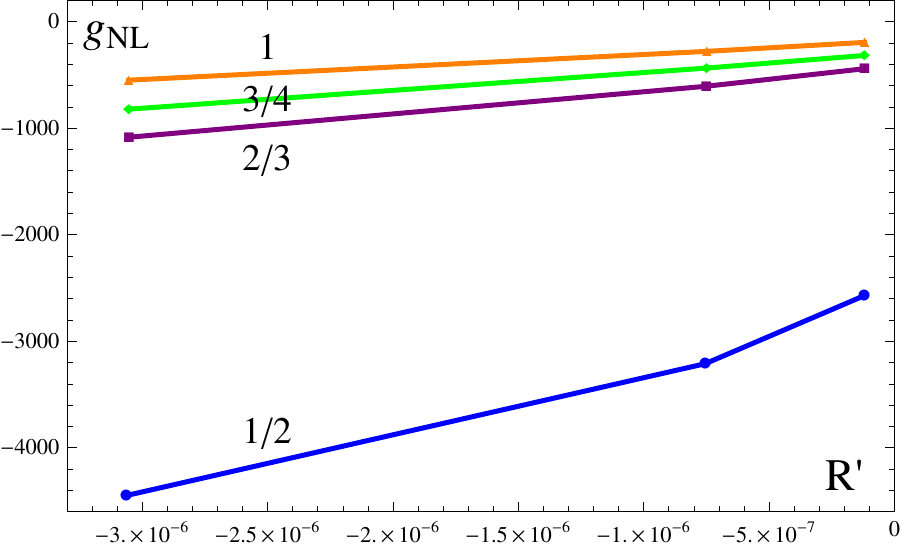}}
\caption{\label{fig:fgNL_Rprime-efolds} Non-Gaussianity plotted against different slopes of the scalar curvature ($R'$) for different durations of the conversion ($N=\sfrac{1}{2}, \sfrac{2}{3}, \sfrac{3}{4}, 1$) for the repulsive potential $V_2$ with $r=1$. The slope $R'$ is varied by choosing different values for $d=\sfrac{1}{10}, \sfrac{1}{20}, \sfrac{1}{50}.$ Note that the magnitudes of $f_{NL}$ and $g_{NL}$ are significantly reduced for smoother conversion processes.}
\end{figure}

\begin{figure}[h!]
\centering
\subfigure[\ $f_{NL}$ vs. $R'$.]
  {\includegraphics[scale=0.85]{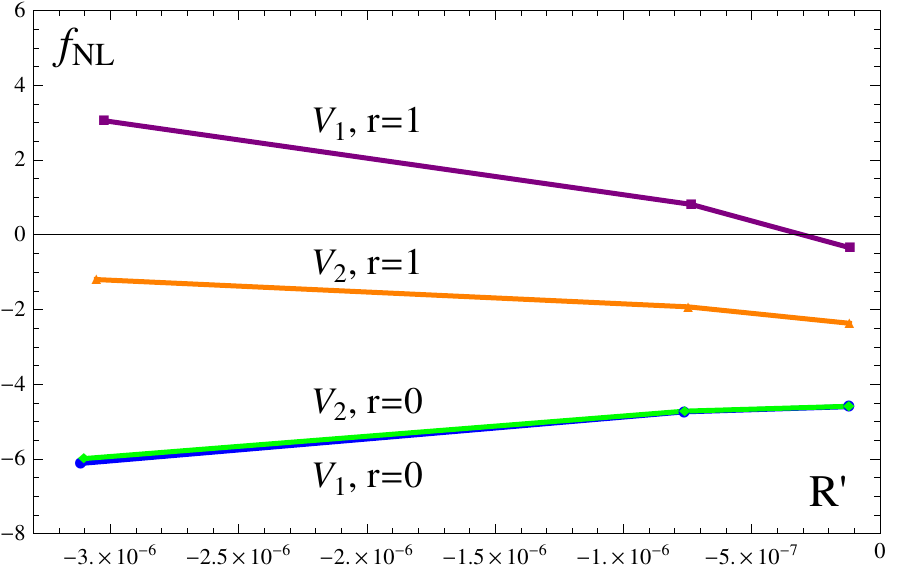}}
\subfigure[\ $g_{NL}$ vs. $R'$.]
  {\includegraphics[scale=0.85]{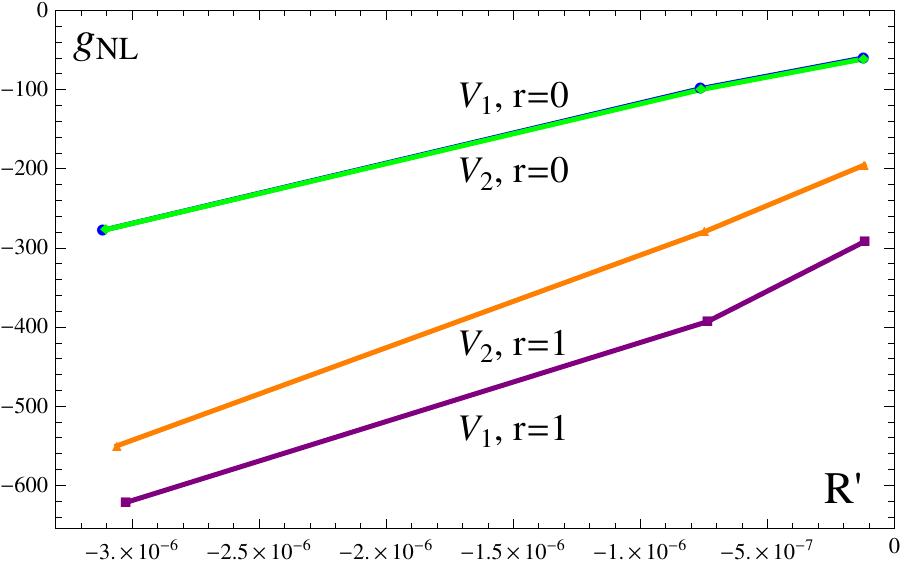}}
\caption{\label{fig:fgNL_Rprime_r} Non-Gaussianity plotted against different slopes of the scalar curvature ($R'$) for different potentials ($V_{1,2}$ with $r = 0, 1$) for a conversion duration of one e-fold. The two $r=0$ lines happen to be virtually coincident. Note that the values for $f_{NL}$ are clustered around zero, while the values for $g_{NL}$ are always appreciably negative. This is a characteristic feature of current ekpyrotic models.}
\end{figure}

Fig.~\ref{fig:fgNL_Rprime-efolds} shows plots of the local non-linearity parameters $f_{NL}$ (parameterising the local bispectrum) and $g_{NL}$ (parameterising the local trispectrum) for different durations of conversion and as a function of the slope $R'$. (The non-linearity parameters of the comoving curvature perturbation are defined in appendix \ref{sec:NG}.) There are two obvious trends: the smoother, and thus the longer and more efficient, the conversion process is, the smaller the non-Gaussianity. And the closer the field space metric is to trivial, again the smaller in magnitude are the non-linearity parameters $f_{NL}$ and $g_{NL}.$ Note that for smoother conversions, the dependence on the slope $R'$ is much weaker, and hence, to some extent, the predictions converge for smooth conversions. Referring back to our previous discussion, it is easy to see by extrapolation that large and rapidly varying field space curvatures very quickly lead to values of the non-Gaussianity parameters that are much larger than current observational bounds allow for. On the other hand, for smooth conversions and small and slowly changing field space curvatures the local bispectrum parameter $f_{NL}$ is of magnitude $|f_{NL}|\lesssim 5$ while the trispectrum parameter is always negative and of magnitude $|g_{NL}| \sim {\cal O}(10^2) - {\cal O}(10^3).$ These values are confirmed by an analysis of the effect of changing the functional form of the repulsive potential (while specialising to smooth conversions lasting one e-fold), as shown in Fig.~\ref{fig:fgNL_Rprime_r}. 
Note that the two potentials $V_1$ and $V_2$ with $r=0$, colour-coded in blue and green, respectively, give nearly identical predictions.

\subsubsection{Asymptotically flat field space metric}

\begin{figure}[h!]
\centering
\subfigure[\label{fig:Exp_phi_chi} Field space trajectory]
  {\includegraphics[scale=0.55]{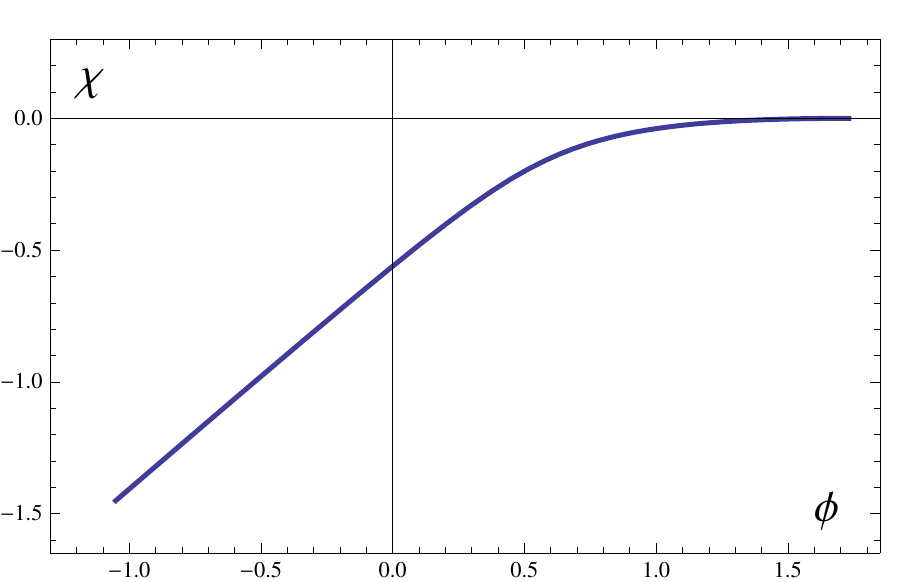}}
\subfigure[\label{fig:Exp_Omega_phi} Field space metric]
  {\includegraphics[scale=0.55]{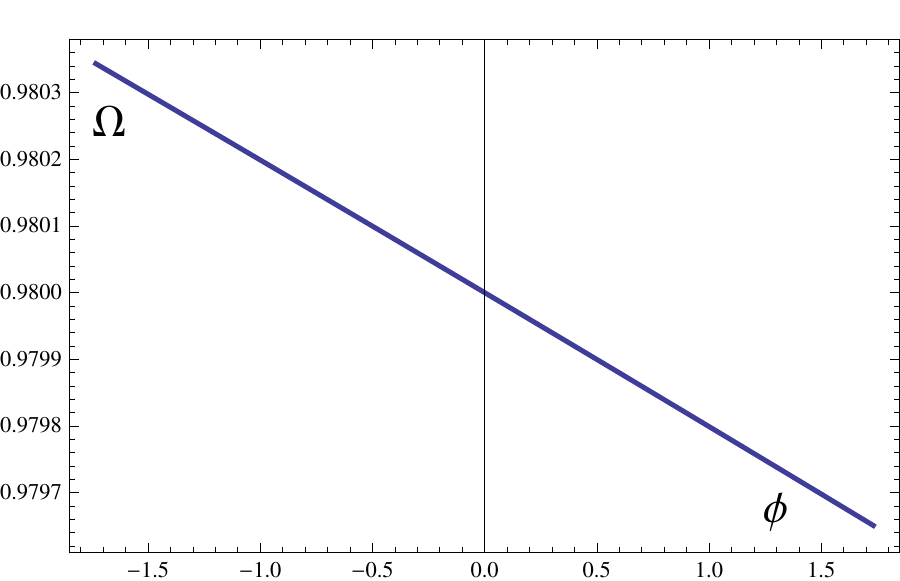}}
\subfigure[\label{fig:Exp_R_t} Field space curvature]
  {\includegraphics[scale=0.538]{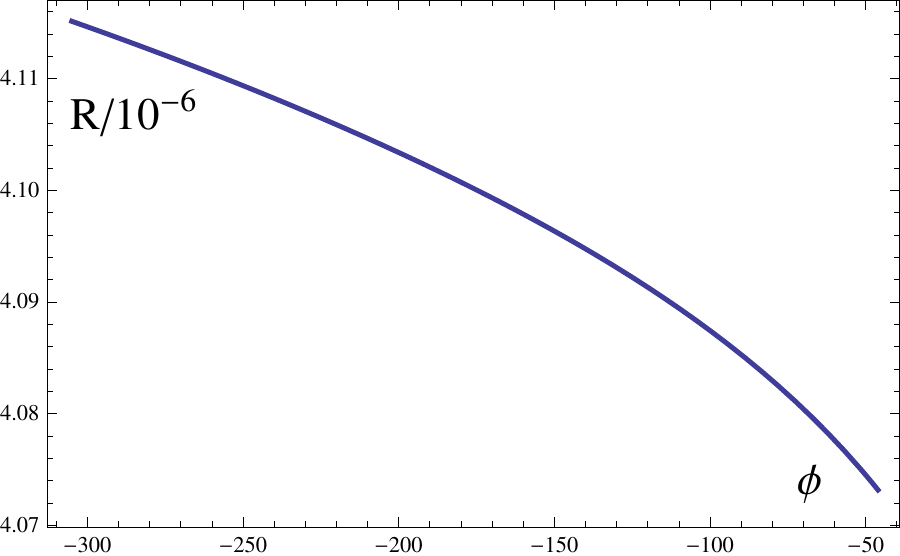}}
\caption{\label{fig:Exp_Omega_phi_chi} The evolution of the fields, the field space metric ($\O = 1 - b \mathrm{e}^{d \f / 2}$) and the scalar curvature during one e-fold of conversion (from $t=-304$ to $t=-46$), plotted for the specified case $r=1$, $b=d=\frac{1}{50}$ giving $f_{\text{NL}}=1.0$ and $g_{\text{NL}}=-235$.}
\end{figure}

In order to check the robustness of our results, we will now consider a different functional form of the metric, namely we will consider the case where a trivial metric is approached exponentially fast (in field space),
\be \label{eq:Omega1-Exp}
\O = 1 - b \mathrm{e}^{d \f / 2},
\ee
where $b$ and $d$ are parameters. We have plotted $\O$ in figure~\ref{fig:Exp_Omega_phi} in the region of the conversion. The corresponding field space trajectory and curvature scalar  are shown in figures \ref{fig:Exp_phi_chi} and \ref{fig:Exp_R_t}, respectively.

\begin{figure}[h!]
\centering
\subfigure[\ $f_{NL}$ vs. $b=d$.]
  {\includegraphics[scale=0.85]{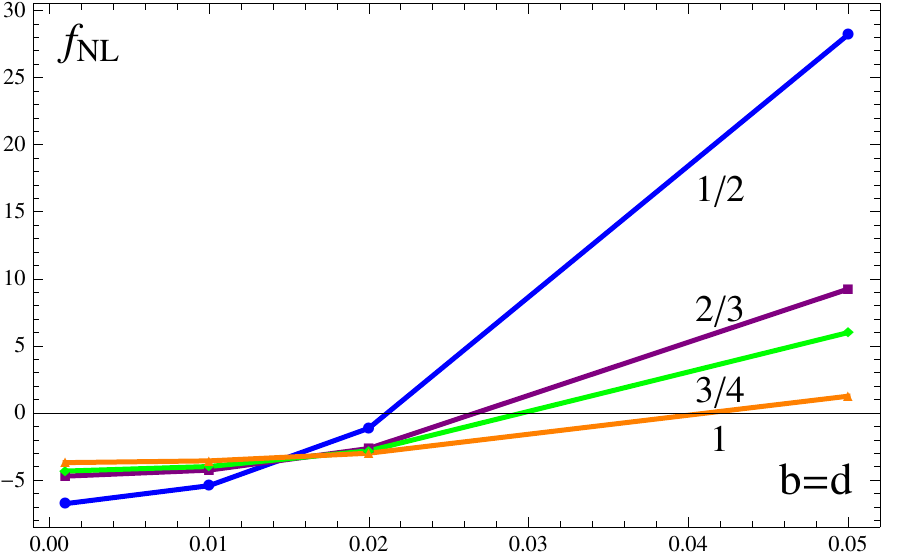}}
\subfigure[\ $g_{NL}$ vs. $b=d$.]
  {\includegraphics[scale=0.85]{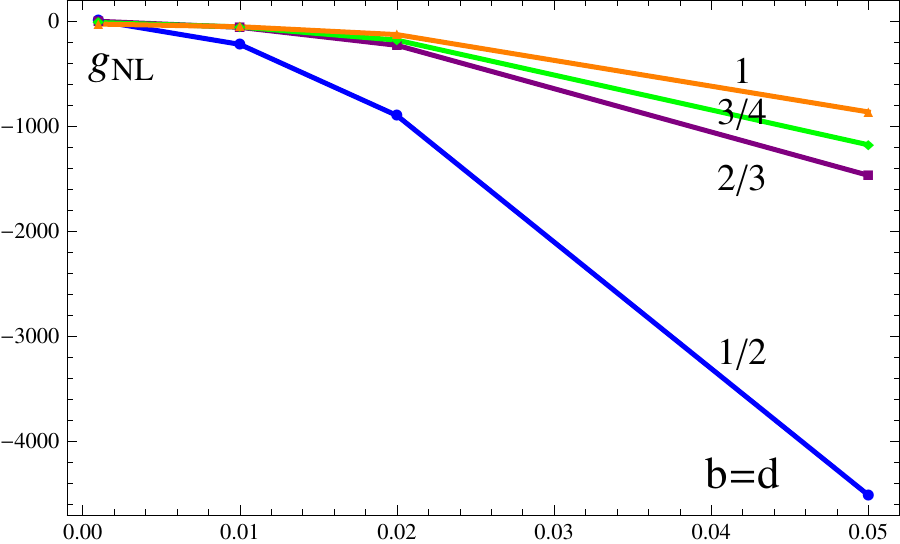}}
\caption{\label{fig:fgNL_bd_efolds} Non-Gaussianity plotted against different field space metrics ($\O = 1 - b \mathrm{e}^{d \f / 2}$ with $b=d$) for different durations of the conversion ($N=\sfrac{1}{2}, \sfrac{2}{3}, \sfrac{3}{4}, 1$) for the potential $V_2$ with $r=0$. Note that as in the case with a linearly decaying field space curvature, the magnitudes of $f_{NL}$ and $g_{NL}$ are significantly reduced for smoother conversion processes.}
\end{figure}

Once again, we can verify the importance of the efficiency of conversion -- see Fig. \ref{fig:fgNL_bd_efolds}. We have plotted the results as a function of $b=d$ -- for $b\neq d$ we found similar results (though typically slightly less variation in the non-linearity parameters). As the figure demonstrates, an efficient/smooth conversion is crucial, in the sense that in this case the typical values of the bispectrum are of ${\cal O}(1).$ Note that for less efficient conversions the spread in values is much larger, and hence no generic predictions can be made. For the trispectrum, the situation is analogous, with efficient conversions drastically reducing the range of possible values of $g_{NL}.$

\begin{figure}[h!]
\centering
\subfigure[\ $f_{NL}$ vs. $b=d$.]
  {\includegraphics[scale=0.85]{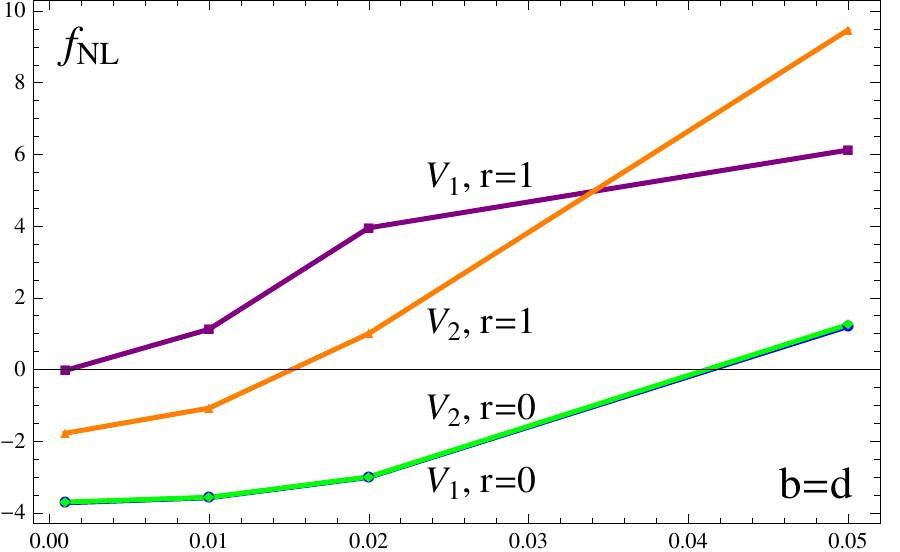}}
\subfigure[\ $g_{NL}$ vs. $b=d$.]
  {\includegraphics[scale=0.85]{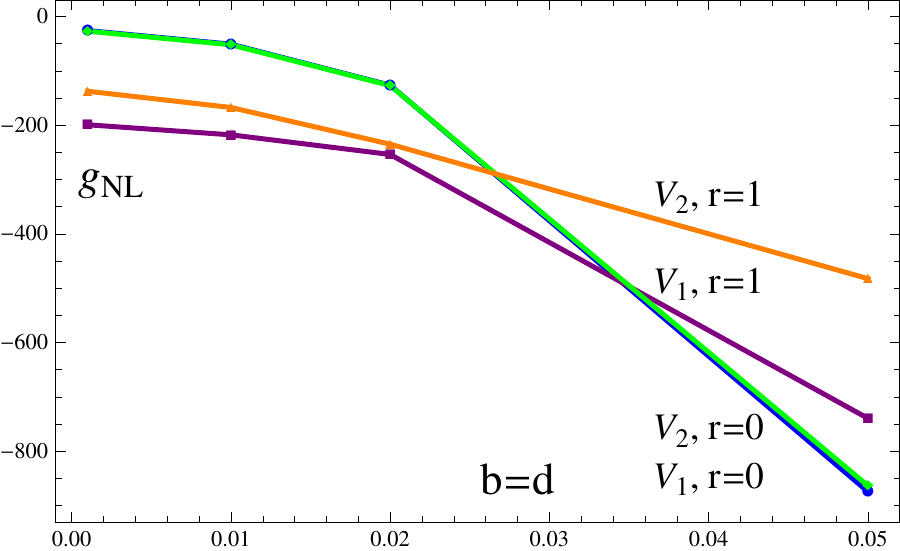}}
\caption{\label{fig:fgNL_bd_r} Non-Gaussianity plotted against different field space metrics ($\O = 1 - b \mathrm{e}^{d \f / 2}$ with $b=d$) for different potentials ($V_{1,2}$ with $r = 0, 1$) for a conversion duration of one e-fold.}
\end{figure}

We can also determine the effect of changing the functional form of the repulsive potential in scalar field space. The results are shown in Fig.~\ref{fig:fgNL_bd_r}, where for all cases we have assumed one e-fold of conversion. As can be seen from the figure, for such smooth conversions the expected values for the bispectrum are in the range $|f_{NL}| \lesssim 5,$ while those for the trispectrum are $|g_{NL}| \sim {\cal O}(10^2) - {\cal O}(10^3)$ and negative in sign, exactly as for the case of a linearly changing scalar field curvature.

\subsection{Comparison to the minimally coupled entropic mechanism}

It may be interesting to compare these results to those obtained via the older, minimally coupled entropic mechanism \cite{Notari:2002yc,Finelli:2002we,Lehners:2007ac}. In that case, the kinetic terms of the scalar fields are canonical, but one assumes a potential that is unstable in the entropic direction. During the ekpyrotic phase, the potential is usefully written as
\be \label{eq:standekppotl}
V_{\text{min. entropic mech., ek}} = - V_0 e^{- \sqrt{2 \ep} \s} \left[ 1 + \ep s^2 + \frac{\kappa_3}{3!} \ep^{3/2} s^3 + \frac{\kappa_4}{4!} \ep^{2} s^4 + \dots  \right],
\ee
where $\kappa_3$ and $\kappa_4$ are important for the bispectrum and trispectrum, respectively. In these models, and in contrast to the non-minimal entropic mechanism, a substantial part of the total non-Gaussianity can already arise during the ekpyrotic phase. This can be seen by solving the equation of motion (\ref{eq:eomdes3_new}) for the entropy perturbation during the ekpyrotic phase. Expanding to leading order in $1/\ep$, for large $\ep$, we have as the initial conditions for the start of the conversion phase
\be \label{eq:standdes}
\de s = \de s_L + \frac{\kappa_3 \sqrt{\ep}}{8} \de s_L^2 + \ep \left( \frac{\kappa_4}{60} +\frac{\kappa_3^2}{80} - \frac{19}{60} \right) \de s_L^3.
\ee
Notice the different numerical factor in the term proportional to $\ep$ compared to \cite{Lehners:2009ja} due to the change in the definition of the third order entropy perturbation. As is clear from this expression, there is typically already a significant non-Gaussian component to the entropy perturbation prior to the phase of conversion. What is more, some of this conversion already occurs during the ekpyrotic phase, where the comoving curvature obeys the evolution equation (where $\z=\z^{(1)}+\z^{(2)}+\z^{(3)}$) 
\be
\z' =   \frac{2 H}{\sb'^2} \left[  - \frac{\Vb_{,\s}}{ \sb'} \de s \de s'   + \Vb_{;ss} \de s^2  + \frac{1}{3} \Vb_{;sss}  \de s^3
\right]\!,
\ee
Using (\ref{eq:zetaNL}), this leads to 
\bea
f_{NL \, \text{integrated}} &=& \frac{5}{12} \frac{\left[ \de s_L (t_{\text{ek-end}}) \right]^2}{|\z_L(t_{\text{conv-end}})|^2},  \\
g_{NL\, \text{integrated}} &=& \frac{275}{1296} \kappa_3 \sqrt{\ep} \frac{\left[ \de s_L (t_{\text{ek-end}}) \right]^3}{|\z_L(t_{\text{conv-end}})|^3}. 
\eea
In order to calculate the contribution from the additional conversion process due to the subsequent bending of the scalar field trajectory, we have solved and integrated the equations of motion (\ref{eq:eomdes3_new}) and (\ref{eq:zeta3comgauge})  numerically, using the expression (\ref{eq:standdes}) as the initial condition for the entropy perturbation.

\begin{figure}
\centering
\subfigure[\label{fig:fNL_kappa3-r} $f_{NL}$ vs. $\kappa_3$.]
  {\includegraphics[scale=0.85]{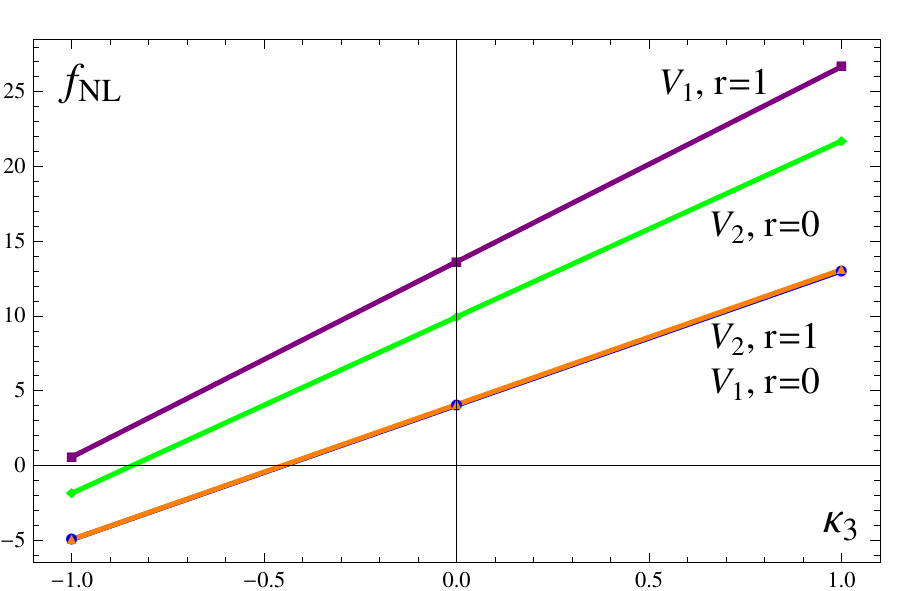}}
\subfigure[\label{fig:gNL_kappa4_kappa30-r} $g_{NL}$ vs. $\kappa_4$ with $\kappa_3=0$.]
  {\includegraphics[scale=0.85]{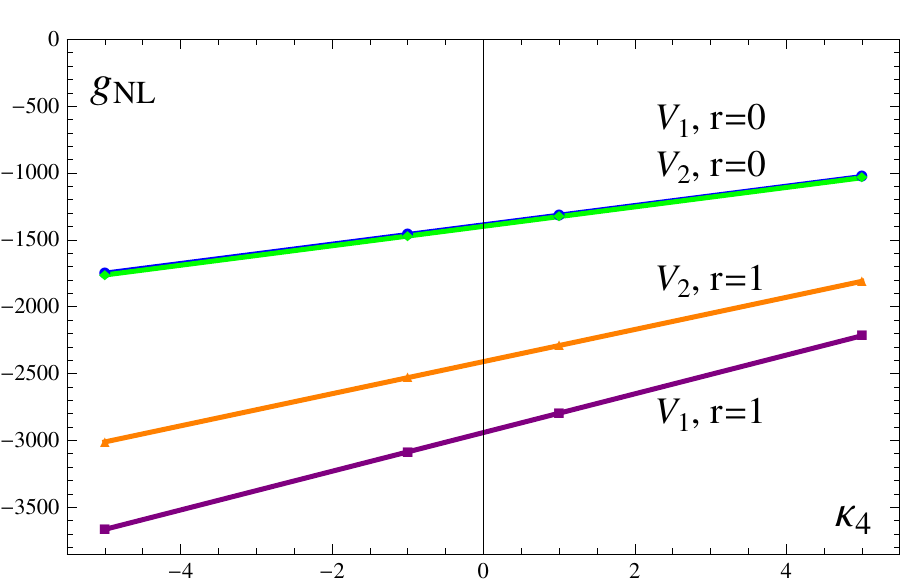}}
\caption{\label{fig:fgNL_kappa34-r} Non-Gaussianity plotted against different potentials ($V_{1,2}$ with $r = 0, 1$) for a conversion duration of one e-fold.}
\end{figure}

The minimal case was analysed in \cite{Lehners:2009ja} in some detail. There it was shown that the range of predicted values for the non-Gaussianity parameters narrows drastically as the conversion process becomes smoother, just as we have found here. Specialising to conversions lasting one e-fold, we have reproduced the results of \cite{Lehners:2009ja}: Fig. \ref{fig:fgNL_kappa34-r} shows the expected values of $f_{NL}$ as a function of $\kappa_3$ and those of $g_{NL}$ as a function of $\kappa_4$ (this time assuming $\kappa_3=0$). As already discussed in \cite{Lehners:2013cka}, one can obtain a bispectrum in agreement with observations by assuming that the potential is (nearly) symmetric, which corresponds to $|\kappa_3| \lesssim 1.$ In this case, the trispectrum remains negative and of ${\cal O}(10^3).$ Thus we see that if we restrict to symmetric potentials, the minimally coupled entropic mechanism leads to similar predictions for the non-Gaussianity parameters $f_{NL}$ and $g_{NL}$ as the non-minimally coupled model considered in the present paper, though $g_{NL}$ is typically up to an order of magnitude larger in the minimally coupled case due to the significant intrinsic contribution represented by the very last ($\kappa_{3,4}$-independent, but $\epsilon$-dependent) term in \eqref{eq:standdes}.

\clearpage
\section{Discussion} \label{section:discussion}
In this paper, we have adopted the covariant formalism to derive exact evolution equations for nonlinear perturbations, in a universe dominated by two scalar fields with a non-trivial field space metric. We have then expanded the equations of motion for the entropy fluctuation (\ref{eq:eomdes3_new}) and the comoving curvature perturbation (\ref{eq:zeta3comgauge}) up to third order in perturbation theory. These equations constitute our main technical result from which the non-linearity parameters for the observed density perturbations can be deduced.

We applied the equations to ekpyrotic models in which the primordial curvature perturbations are generated via the non-minimal entropic mechanism. In these models, in a two-stage process nearly scale-invariant entropy perturbations are generated first due to the non-minimal kinetic coupling between two scalar fields. Subsequently, these perturbations are converted into curvature perturbations by a bending in the field space trajectory. Solving the equations of motion analytically during the ekpyrotic phase we find vanishing bi- and trispectra for the entropy perturbations. However, this property is significantly modified during the conversion process to curvature perturbations.

Indeed, what we find is that the efficiency of the conversion process is crucial: inefficient conversions would lead to curvature perturbations with a small amplitude and very large and wildly varying non-Gaussianities. On the other hand, for efficient conversions the results converge and lead to the following predictions (which we compare to the current observational bounds \cite{Ade:2015ava}):
\begin{eqnarray}
&\textrm{Non-minimal entropic mechanism                 } \qquad & \textrm{Observational bounds} \nonumber \\ & |f_{NL}^{\text{local}}| \lesssim 5 & f_{NL}^{\text{local}} = 0.8 \pm 5.0 \quad(1\s) \\ & g_{NL}^{\text{local}} \sim {\cal O}(-10^2) \textrm{ or } {\cal O}(-10^3) & g_{NL}^{\text{local}} = (\, 9.0 \pm 7.7) \times 10^4 \quad(1\s) \\ & \alpha_s \sim {\cal O}(-10^{-3}) & \al_s = \, 0.003 \pm 0.007 \quad(1\s)
\end{eqnarray}
Here, for completeness, we have added the prediction for the running of the spectral index $\alpha_s \equiv \frac{\d n_s}{\d \ln k}$ that is expected in these models \cite{Lehners:2015mra}. Note the highly interesting prediction that all three observables ought to actually be observable in the near future. Also, an important feature is that $f_{NL}$ may be small, but $g_{NL}$ is typically not simultaneously close to zero as well, and in fact there is a clear correlation between all observables, as both the running and $g_{NL}$ are expected to be negative and significant. As in all currently known ekpyrotic models, one would not expect to see any primordial gravitational waves. Thus the present model has the potential to be refuted or supported by observations with significant levels of confidence.

As a final comment, let us return to the issue of the efficiency of conversion and model building. As we saw, the kinetic coupling between the two scalar fields has to return to trivial after the ekpyrotic phase, in much the same way as the potential has to turn off. One may wonder whether such a feature could arise in a plausible manner from the point of view of fundamental physics. A more complete answer to this question will of course have to await further developments in fundamental physics, and especially in quantum gravity, but we would like to exhibit one example where such a feature is indeed seen. This comes from considering supergravity coupled to scalar fields with higher-derivative kinetic terms \cite{Khoury:2010gb,Koehn:2012ar}. In this class of models, the higher-derivative terms add corrections to both the ordinary kinetic terms and the potential of the theory, even when the higher-derivative terms are not significant dynamically. More precisely, in these theories the bossing contribution is of the form
\begin{equation}
(\partial A)^2 (\partial A^\star)^2 - 2 \, e^{K/3} F F^\star \, \partial A \cdot \partial A^\star + e^{2K/3} (F F^\star)^2,
\end{equation}
where $A$ is a complex scalar field (or may be thought of as two real scalars, just as in the theories considered here), while $F$ is a complex auxiliary field and $K$ is the K\"{a}her potential, which is just a function of $A$ and $A^\star$. The value of the auxiliary field depends on the superpotential -- crucially, $F$ is small when the superpotential is small. Now keeping in mind that the expression above is a correction term to the usual kinetic terms, we see that when the superpotential becomes unimportant, then the potential in the theory turns off but so does the correction to the kinetic term $FF^\star \, \partial A \cdot \partial A^\star.$ This is exactly what would be required for the conversion process in the class of models we have analysed in the present work. It would certainly be interesting to see whether a more complete embedding in supergravity can be realised. This is a topic we will leave for future work.


\acknowledgements 
We gratefully acknowledge the support of the European Research Council in the form of the Starting Grant No. 256994 ``StringCosmOS''.

\appendix

\section{Local non-linearity parameters} \label{sec:NG}
The observable that is relevant for comparison with observations is the comoving curvature perturbation,
$\z = \z^{(1)} + \z^{(2)} + \z^{(3)} + \cdots$. Linear (Gaussian) perturbations are related to observations of the power spectrum, $P(k_1)$, defined by the 2-point correlation function, 
\be
\left\langle \z_{k_1} \z_{k_2}  \right\rangle = (2 \pi)^3 \de^3 \left(\bf{k_1} + \bf{k_2} \right) P(k_1).
\ee
Similarly, quadratic and cubic corrections to these perturbations are related to observations of the 3- and 4-point functions, respectively. For an exactly Gaussian probability distribution all information is contained in the 2-point correlation function. In particular this implies that for odd $n$, all $n$-point functions are zero, while for even $n$ the $n$-point functions can be written as products of 2-point functions. The bispectrum, i.e. the 3-point correlation function, is defined as
\be
\left\langle \z_{k_1} \z_{k_2} \z_{k_3} \right\rangle = (2 \pi)^3 \de^3 \left(\bf{k_1} + \bf{k_2} + \bf{k_3}  \right) B(k_1, k_2, k_3).
\ee
The connected part of the 4-point function which is not already captured by the product of two 2-point functions is given by the trispectrum,
\be
\left\langle \z_{k_1} \z_{k_2} \z_{k_3} \z_{k_4} \right\rangle = (2 \pi)^3 \de^3 \left(\bf{k_1} + \bf{k_2} + \bf{k_3} + \bf{k_4} \right) T(k_1, k_2, k_3, k_4).
\ee
The $\de$-functions result from momentum conservation, while $B$ and $T$ are shape functions for a closed triangle and a quadrangle, respectively.

In momentum space, B is then parameterised by the shape function $f_{NL}$, via
\be
B = \frac{6}{5} f_{NL} \left[P(k_1) P(k_2)+2 \,\text{permutations} \right].
\ee
$T$ describes two different shape functions parameterised by $\tau_{NL}$ and $g_{NL}$, see e.g. \cite{Byrnes:2006vq} for additional details. These are defined by
\be
T = \tau_{NL} \left[P(k_{13})P(k_3)P(k_4)+11 \,\text{permutations} \right]+ \frac{54}{25} g_{NL}\left[P(k_2)P(k_3)P(k_4) + 3 \,\text{permutations} \right]\!,
\ee
where $\bf{k_{ij}} = \bf{k_{i}} + \bf{k_{j}}$.

For the local types of non-Gaussianity that are relevant for the models we consider, the parameters $f_{NL}$ and $g_{NL}$ can also be related to the (real space) expansion of the curvature perturbation on uniform energy density surfaces in terms of its Gaussian component $\z_L$, via
\be\label{eq:zetaNL}
\z = \z_L + \frac{3}{5} f_{NL} \z_L^2 + \frac{9}{25} g_{NL} \z_L^3,
\ee
which is related to the Bardeen space-space metric perturbation $\Phi_H = \Phi_L +f_{NL} \Phi_L^2 +g_{NL} \Phi_L^3$ \cite{Bardeen:1980kt} 
through $\z_L = \frac{5}{3} \Phi_L$ during the era of matter domination. For models in which the density perturbations originate from the dynamics of a single field (as in our case, where the origin of the perturbations is originally solely from the entropy field) $\tau_{NL}$ is directly related to the square of $f_{NL}$ -- explicitly we have
\be
\tau_{NL} = \frac{36}{25} f_{NL}^2.
\ee
In order to obtain the non-Gaussianity parameters we first have to solve the equation of motion for $\de s$ up to third order in perturbation theory.
This allows us to integrate the equation of motion for $\z'$, also at the first three orders in perturbation theory. We then obtain the local non-linearity parameters by evaluating
\bea
f_{NL} &=& \frac{5}{3} \frac{\int_{t_{\text{ekp-beg}}}^{t_{\text{conv-end}}} \z^{(2)'}}{\left(\int_{t_{\text{ekp-beg}}}^{t_{\text{conv-end}}} \z^{(1)'}\right)^2}, \\
g_{NL} &=& \frac{25}{9} \frac{\int_{t_{\text{ekp-beg}}}^{t_{\text{conv-end}}} \z^{(3)'}}{\left(\int_{t_{\text{ekp-beg}}}^{t_{\text{conv-end}}} \z^{(1)'}\right)^3},
\eea
where the integrals are evaluated from the time $t_{\text{ekp-beg}}$ that the ekpyrotic phase begins until the conversion process has been completed at $t_{\text{conv-end}}$ and $\z$ has evolved to a constant value.

\section{A new definition of the entropy perturbation $\de s^{(3)}$} \label{sec:O3T0}
I this section of the appendix we show how we are led to defining $\de s^{(3)}$ as given in equation (\ref{eq:des3}).
Expanding the equation of motion for $s_a$ (\ref{eq:eomsa}) to third order without the extra term, i.e. with $\de s^{(3)}|_{\text{old}}=\de s^{(3)} - \frac{1}{6 \sb'^2} \de s \de s'^2$, we obtain
\be \label{eq:eomdes3}
\ba
&0 \approx \de s^{(3)}|_{\text{old}}'' + 3H \de s^{(3)}|_{\text{old}}' + \left( \bar{V}_{;ss} + 3 \tb'^2 + \sb'^2 \bar{e}_{s}^{I} \bar{e}_{s}^{J} \bar{e}_{\s}^{K} \bar{e}_{\s}^{L}   \bar{R}_{IKJL} \right)  \de s^{(3)}|_{\text{old}}  + 2 \frac{\tb'}{\sb'} \de s' \de s^{(2)'} \\
&+\left( \frac{2}{\sb'} \tb'' +\frac{2}{\sb'^2} \Vb_{,\s} \tb' - \frac{3}{\sb'} H \tb' \right) \left( \de s \de s^{(2)} \right)' \\
&+\left( \Vb_{;sss} - \frac{10}{\sb'} \Vb_{;ss} \tb'  - \frac{18}{\sb'} \tb'^3 +  \ebs^I \ebs^J \ebsi^K \ebsi^L  \left( -2 \sb' \tb' \Rb_{IKJL} + \sb'^2 \ebs^N \cD_N \Rb_{IKJL}  \right) \right) \de s \de s^{(2)} \\
&+ \frac{\Vb_{,\s}}{\sb'^3} \de s'^3 
+ \frac{1}{\sb'^2} \left[\Vb_{;\s\s} + \frac{3 \Vb_{,\s}^2}{\sb'^2} + \frac{3}{\sb'} H \Vb_{,\s} - 2 \Vb_{;ss} - 6\tb'^2 - \bar{e}_{s}^{I} \bar{e}_{s}^{J} \bar{e}_{\s}^{K} \bar{e}_{\s}^{L}  \bar{R}_{IKJL}  \right] \de s \de s'^2 \\
&+ \left[ - \frac{10}{\sb'^2} \tb' \tb'' - \frac{3}{2 \sb'} \Vb_{;ss\s} 
- \frac{5}{\sb'^3} \Vb_{;ss} \Vb_{,\s} - \frac{7}{\sb'^3} \Vb_{,\s} \tb'^2 - \frac{3}{\sb'^2} H \Vb_{;ss} +\frac{14}{\sb'^2} H \tb'^2 \parenthnewln 
-2 \frac{ \Vb_{,\s} }{\sb'}  \bar{e}_{s}^{I} \bar{e}_{s}^{J} \bar{e}_{\s}^{K} \bar{e}_{\s}^{L}  \bar{R}_{IKJL} 
- \frac{\sb' }{2} \bar{e}_{s}^{I} \bar{e}_{s}^{J} \bar{e}_{\s}^{K} \bar{e}_{\s}^{L} \bar{e}_{\s}^{M} \cD_M  \bar{R}_{IKJL}   \right] \de s^2 \de s' \\
&+ \bigg[ \frac{1}{6} \Vb_{;ssss} - \frac{7}{3 \sb'} \Vb_{;sss} \tb' + \frac{2}{\sb'^2} \Vb_{;ss}^2 + \frac{21}{\sb'^2} \Vb_{;ss} \tb'^2 + \frac{27}{\sb'^2} \tb'^4 
+ \frac{\sb'^2}{3} \left( \ebs^I \ebs^J \ebsi^K \ebsi^L \Rb_{IKJL}  \right)^2
\parenthnewlnbigg
+ \frac{1}{3} \ebs^I \ebs^J \ebsi^K \ebsi^L \left( \Rb_{IKJL}  \left( 3 \Vb_{;ss} + 7 \tb'^2 \right) 
-2 \sb' \tb'  \ebs^N \cD_N \Rb_{IKJL}  
\parenthnewlnlnbigg
+ \sb'^2 \ebs^N \ebs^Q \left(   \h \cD_Q \cD_N \bar{R}_{IKJL} -  \Rb_{IKJP} \Rb^P_{\; NLQ} + \Rb_{IKJL} \Rb^P_{\; NPQ} \right) 
\right)\bigg] \de s^3,
\ea
\ee
where we have used $V_i \approx 0$ and $V^{(3)}_i \approx 0$ on large scales.
This equation reduces to the one derived in \cite{Lehners:2009ja} for a trivial field space metric. During the non-minimally coupled ekpyrotic phase the equation of motion for $\de s^{(3)}$ simplifies to
\be
\ba \label{eq:eomdes3ekp}
\de s^{(3)}|_{\text{old}}'' &+ 3 H \de s^{(3)}|_{\text{old}}' + \left[ \O^{-1} \O_{,\f} \bar{V}_{,\f}  - \O^{-1} \O_{,\f\f} \bar{\f}'^2 \right]  \de s^{(3)}|_{\text{old}}  \\
&+ \fb'^2  \left[ \frac{2}{3} \O^{-3} \O_{,\f}^2 \O_{,\f\f} +\frac{2}{3} \O^{-2} \O_{,\f\f}^2 +\frac{1}{3} \O^{-2} \O_{,\f} \O_{,\f\f\f} - \frac{1}{3 \fb'^2} \O^{-2} \O_{,\f} \O_{,\f\f} \Vb_{,\f} \right] \de s^3
 \approx 0,
\ea
\ee
where we used $\de s^{(2)}=0$ as well as $\de s' =- \sb' \O^{-1} \O_{,\f} \de s$.
However, the third order entropy perturbation, given by
\be \label{eq:des3ekp}
\de s^{(3)} |_{\text{old,ekp}} = - \O \de \chi^{(3)} - \frac{1}{6} \O^{-2} \O_{,\f}^2 \de s^3,
\ee
contains a $\de s^3$-term acting as a source in the equation of motion, while $\de \chi^{(3)} =0$ trivially solves the equation of motion (\ref{eq:eomdes3ekp}). It is certainly more natural to define the entropic perturbation in such a way that during the non-minimal ekpyrotic phase the solution is given by $\de s^{(3)}=\de \chi^{(3)}=0.$ This motivates us to add a gauge-invariant term to the definition of $\de s^{(3)}$ that reduces to $+\frac{1}{6} \O^{-2} \O_{,\f}^2 \de s^3$ during the ekpyrotic phase. The choice is not unique however, and the terms that can be added are
\be
T = A \de s'^3 + B \de s \de s'^2 + C \de s^2 \de s' + D \de s^3,
\ee
where we leave $A,B,C,D$ arbitrary for now. During the ekpyrotic phase, $\de s' = - \sb' \O^{-1} \O_{,\f} \de s$, and hence we need
\be \label{eq:Tekp}
T = \frac{1}{6} \O^{-2} \O_{,\f}^2 \de s^3 =\left[ - A \sb'^3 \O^{-3} \O_{,\f}^3 + B \sb'^2 \O^{-2} \O_{,\f}^2 - C \sb' \O^{-1} \O_{,\f} + D \right] \de s^3.
\ee
It is immediately clear that we can set $A=0$. The derivative of the term $T$ added to the definition of $\de s^{(3)}$ has to be subtracted off $\de s_i^{(3)}$, giving the following contribution to the entropic equation of motion at third order: 
\be
\ba
-& \left[ T'' + 3H T' + \left( \bar{V}_{;ss} + 3 \tb'^2 + \sb'^2 \bar{e}_{s}^{I} \bar{e}_{s}^{J} \bar{e}_{\s}^{K} \bar{e}_{\s}^{L}   \bar{R}_{IKJL} \right) T \right] \\
&=\de s^3 \bigg[ - D'' + 2 \left[  D +  C' - B \left( \bar{V}_{;ss} + 3 \tb'^2 + \sb'^2 \bar{e}_{s}^{I} \bar{e}_{s}^{J} \bar{e}_{\s}^{K} \bar{e}_{\s}^{L}   \bar{R}_{IKJL} \right)\right]  \parenthnewlnbigg
\left( \bar{V}_{;ss} + 3 \tb'^2 + \sb'^2 \bar{e}_{s}^{I} \bar{e}_{s}^{J} \bar{e}_{\s}^{K} \bar{e}_{\s}^{L}   \bar{R}_{IKJL} \right) 
- 3 H D' + C\left[ \sb' \Vb_{;ss\s} - 6 H \tb'^2 - 4 \frac{\tb'^2}{\sb'} \Vb_{,\s} + 8 \tb' \tb'' \parenthnewlnlnbigg
- 2 \sb' \left( 3 H \sb' + \Vb_{,\s}  \right)  \bar{e}_{s}^{I} \bar{e}_{s}^{J} \bar{e}_{\s}^{K} \bar{e}_{\s}^{L}   \bar{R}_{IKJL} + \sb'^3  \bar{e}_{s}^{I} \bar{e}_{s}^{J} \bar{e}_{\s}^{K} \bar{e}_{\s}^{L} \bar{e}_{\s}^{M} \cD_M  \bar{R}_{IKJL} \right] \bigg] \\
&+ \de s^2 \de s' \bigg[-6 D' - C'' + 3 H' C + 3 H C' +\left( 6 C + 4 B' -12 H B \right) \parenthnewlnbigg
\left( \bar{V}_{;ss} + 3 \tb'^2 + \sb'^2 \bar{e}_{s}^{I} \bar{e}_{s}^{J} \bar{e}_{\s}^{K} \bar{e}_{\s}^{L}   \bar{R}_{IKJL} \right)  
 + 2B \left[ \sb' \Vb_{;ss\s} - 6 H \tb'^2 - 4 \frac{\tb'^2}{\sb'} \Vb_{,\s} + 8 \tb' \tb'' \parenthnewlnlnbigg
- 2 \sb' \left( 3 H \sb' + \Vb_{,\s}  \right)  \bar{e}_{s}^{I} \bar{e}_{s}^{J} \bar{e}_{\s}^{K} \bar{e}_{\s}^{L}   \bar{R}_{IKJL} + \sb'^3  \bar{e}_{s}^{I} \bar{e}_{s}^{J} \bar{e}_{\s}^{K} \bar{e}_{\s}^{L} \bar{e}_{\s}^{M} \cD_M  \bar{R}_{IKJL} \right] \bigg] \\
&+ \de s \de s'^2 \left[-6 D - 4 C' + 3H \left( 4 C - 6 H B \right) - B'' + 6 H' B + 9 H B'  \parenthnewln
+6 B \left( \bar{V}_{;ss} + 3 \tb'^2 + \sb'^2 \bar{e}_{s}^{I} \bar{e}_{s}^{J} \bar{e}_{\s}^{K} \bar{e}_{\s}^{L}   \bar{R}_{IKJL} \right)  \right] \\
&+\de s'^3 \left[-2 C - 2 B' + 12H B \right] .
\ea
\ee
In order to satisfy equation (\ref{eq:Tekp}) during the ekpyrotic phase $C$ would have to contain one Christoffel symbol, and $D$ the product of two $\Gb$'s. As can be seen from the previous equation, this is problematic as the terms that would be added to the equation of motion are not covariant. Take the simple example of $C \sim \Gb$: there are no terms in the equation of motion that can be combined with the new term $\sim \Gb \de s'^3$ to make it covariant. Similarly for $D \neq 0$.
We are forced to choose $B=\frac{1}{6 \sb'^2}$ with $C=D=0$.

\section{Useful formulae} \label{section:appendixB}

\textbf{Metric:} \label{app:metric}
On large scales where spatial gradients can be neglected, the metric can be written as
\be
\text{d}s^2 = - \left(1 + 2 A \right) \text{d} t^2 + a(t)^2 \left(1 - 2 \psi \right) \de_{ij} \text{d}x^i \text{d}x_i,
\ee
where $A= A^{(1)}+A^{(2)}+A^{(3)}$, and $\psi= \psi^{(1)}+\psi^{(2)}+\psi^{(3)}$ up to third order.

Thus, the $00$-component of the inverse metric is given by
\be
g^{00} = -1 + 2A^{(1)} + 2A^{(2)} - 4\left(A^{(1)}\right)^2 +2A^{(3)} - 8 A^{(1)} A^{(2)} + 8 \left(A^{(1)}\right)^3 = - u^0 u^0,
\ee
from which we can deduce
\be \label{eq:u0}
u^0 = 1 - A^{(1)} - A^{(2)} + \frac{3}{2} \left(A^{(1)}\right)^2 - A^{(3)} +3 A^{(1)} A^{(2)} - \frac{5}{2} \left(A^{(1)}\right)^3.
\ee
Moreover, for simplicity we choose $u^{a}$ such that $u_i = 0$, and on large scales we can show that $u^i \approx 0$.

\textbf{Scalar field perturbations:}
Rewriting the perturbation in the scalar fields in terms of adiabatic and entropic fields we have
\be \label{phiJ}
\de \phi^J = \ebsi^J \de \s + \ebs^J \de s,
\ee
\be\label{phiJ'}
\de \phi^{J'} = \ebsi^J \de \s' + \ebs^J \de s' + \tb' \left( \ebs^J \de \s - \ebsi^J \de s \right)
- \sb' \Gb_{LK}^J \ebsi^L \left( \ebsi^K \de \s + \ebs^K \de s \right),
\ee
at linear order and
\be \label{phi2J}
\ba
\de \phi^{(2)J} &=\ebsi^J \left[\de \s^{(2)} - \frac{1}{2 \sb'} \de s \de s' \right] 
+ \ebs^J \left[ \de s^{(2)} + \frac{\de \s}{\sb'} \left( \de s' +\frac{\tb'}{2} \de \s \right) \right]  \\
&- \h \Gb^J_{LK} \left( \ebsi^L \de \s + \ebs^L \de s \right) \left( \ebsi^K \de \s + \ebs^K \de s \right),
\ea
\ee
\be \label{phi2J'}
\ba
\de \phi^{(2)J'} &= \ebsi^J \left[\de \s^{(2)} - \frac{1}{2 \sb'} \de s \de s'  \right]' 
+ \ebs^J \left[ \de s^{(2)} + \frac{\de \s}{\sb'} \left( \de s' +\frac{\tb'}{2} \de \s \right) \right]' \\
&+ \tb' \left[ \ebs^J \left(\de \s^{(2)} - \frac{1}{2 \sb'} \de s \de s' \right) - \ebsi^J \left( \de s^{(2)} + \frac{\de \s}{\sb'} \left( \de s' +\frac{\tb'}{2} \de \s \right) \right) \right] \\ 
&- \sb' \Gb_{MN}^J \ebsi^M \left[ \ebsi^N \left(\de \s^{(2)} - \frac{1}{2 \sb'} \de s \de s' \right) + \ebs^N \left( \de s^{(2)} + \frac{\de \s}{\sb'} \left( \de s' +\frac{\tb'}{2} \de \s \right) \right)  \right] \\
&- \Gb_{LK}^J \left( \ebsi^L \de \s + \ebs^L \de s \right) \left[ \ebsi^K \left(\de \s' - \tb' \de s \right) + \ebs^K \left( \de s' + \tb' \de \s \right) \right] \\
&+  \sb' \ebsi^M \left( \ebsi^L \de \s + \ebs^L \de s  \right) \left( \ebsi^K \de \s + \ebs^K \de s  \right)
 \left[ - \h \p_M \Gb_{LK}^J +  \Gb^J_{LN} \Gb_{MK}^N \right]
\ea
\ee
at second order. The perturbation in the scalar field at third order in \textit{comoving gauge} is given by\footnote{This includes the term $T=\frac{1}{6 \sb'^2} \de s \de s'^2$ from the new defintion of $\de s^{(3)}$.}
\be \label{phi3J}
\ba
\de \phi^{(3)J} &\approx \ebs^J \left[ \de s^{(3)}  - \frac{1}{6 \sb'^2} \de s \de s'^2 \right] - \ebsi^J \left[\frac{1}{2 \sb'} \left(\de s \de s^{(2)'} + \de s' \de s^{(2)} \right) + \frac{\tb'}{6 \sb'^2} \de s^2 \de s' \right]  \\
&- \Gb^J_{KL}  \ebs^L \de s \left[ \ebs^K \de s^{(2)} - \ebsi^K \frac{1}{2 \sb'} \de s \de s' \right]
-\frac{1}{6} \ebs^K \ebs^L \ebs^I \left[ \p_I \Gb^J_{KL} - 2 \Gb^J_{IP} \Gb^P_{KL} \right] \de s^3.
\ea
\ee

\textbf{Field space metric:}
Explicitely, we have (using equations (\ref{phiJ}) and (\ref{phi2J}))
\be
\ba
\de G_{IJ} &= \bar{G}_{IJ,K} \left(\ebsi^K \de \s + \ebs^K \de s \right)
\ea
\ee
at linear order, and
\be
\ba
\de G_{IJ}^{(2)} &= \bar{G}_{IJ,K} \left[ \ebsi^K \left(\de \s^{(2)} - \frac{1}{2 \sb'} \de s \de s' \right) 
+ \ebs^K \left( \de s^{(2)} + \frac{\de \s}{\sb'} \left( \de s' +\frac{\tb'}{2} \de \s \right) \right) \parenthnewln
- \h \Gb^K_{MN} \left( \ebsi^M \de \s + \ebs^M \de s \right) \left( \ebsi^N \de \s + \ebs^N \de s \right) 
\right] \\
&+ \h \bar{G}_{IJ,KL} \left(\ebsi^K \de \s + \ebs^K \de s \right) \left(\ebsi^L \de \s + \ebs^L \de s \right)
\ea
\ee
at second order.

\textbf{Riemann tensor:}
The Riemann tensor with all indices downstairs is given by
\be
\ba
\Rb_{IKJL} &= \bar{G}_{IM} \Rb^M_{\;\; KJL} = \bar{G}_{IM} \left( \p_J \Gb^M_{\; LK} - \p_L \Gb^M_{\; JK} + \Gb^M_{\; JP} \Gb^P_{\; LK} - \Gb^M_{\; LP} \Gb^P_{\; JK} \right) \\
&= \h \left( \bar{G}_{JK,IL} - \bar{G}_{KL,IJ} - \bar{G}_{IJ,KL} + \bar{G}_{IL,KJ} \right) +  \Gb_{\; PIL} \Gb^P_{\; JK} - \Gb_{\; PIJ} \Gb^P_{\; LK},
\ea
\ee
where the Christoffel symbol with all indices downstairs is defined as
\be
\Gb_{IJK} \equiv \bar{G}_{IP} \Gb^P_{JK} = \h \left( \bar{G}_{IJ,K} + \bar{G}_{IK,J} - \bar{G}_{JK,I} \right).
\ee

\textbf{Vielbeine:} 
Expanding the $\s$-vielbein, $e_{\s}^J \equiv \frac{\dot{\phi}^J}{\dot{\s}}$, up to second order, we obtain at linear order
\be \label{eq:deesigmaJ}
\ba
\de e_{\s}^J &= \sb'^{-1} \left( \de s' + \tb' \de \s \right) \ebs^J - \Gb^J_{KL} \ebsi^L \left( \ebsi^K \de \s + \ebs^K \de s \right),
\ea
\ee
and with field space index lowered
\be \label{eq:deesigma_I}
\ba
\de e_{\s I} &= \sb'^{-1} \left( \de s' + \tb' \de \s \right) \eb_{s I} + \Gb_{LKI} \ebsi^L \left( \ebsi^K \de \s + \ebs^K \de s \right).
\ea
\ee
At second order we have
\be\label{eq:de2esigmaJ}
\ba
\de e_{\s}^{(2) J} 
&= - \frac{1}{2 \sb'^2} \ebsi^J \left( \de s' + \tb' \de \s \right)^2 + \frac{1}{\sb'} \ebs^J\bigg[ - \sb'^{-1} \left( \de \s' - \tb' \de s \right) \left( \de s' + \tb' \de \s \right) \parenthnewlnbigg
+ \left( \de s^{(2)} + \frac{\de \s}{\sb'} \left( \de s' +\frac{\tb'}{2} \de \s \right) \right)' 
+ \tb' \left(\de \s^{(2)} - \frac{1}{2 \sb'} \de s \de s' \right)  \bigg] \\
&+ \h \ebsi^J  \ebs^M \ebs^N \ebsi^K \ebsi^L \Rb_{MKNL} \de s^2
- \Gb^J_{KL} \left[\ebsi^L \left[ \ebsi^K \left(\de \s^{(2)} - \frac{1}{2 \sb'} \de s \de s' \right) \parenthnewlnln
+ \ebs^K \left( \de s^{(2)} + \frac{\de \s}{\sb'} \left( \de s' +\frac{\tb'}{2} \de \s \right) \right) \right]
+ \frac{1}{\sb'} \ebs^L \left(\ebsi^K \de \s + \ebs^K \de s \right) \left( \de s' + \tb' \de \s \right) \right]\\
&+ \ebsi^M \left(\ebsi^K \de \s + \ebs^K \de s \right) \left(\ebsi^L \de \s + \ebs^L \de s \right) \left[ - \h \p_M \Gb^J_{KL} + \Gb^J_{LN} \Gb^N_{MK} \right],
\ea
\ee
and
\be\label{eq:de2esigma_I}
\ba
\de e_{\s I}^{(2)} 
&=- \frac{1}{2 \sb'^2} \eb_{\s I} \left( \de s' + \tb' \de \s \right)^2 + \frac{1}{\sb'} \eb_{s I} \bigg[ - \sb'^{-1} \left( \de \s' - \tb' \de s \right) \left( \de s' + \tb' \de \s \right) \parenthnewlnbigg
+ \left( \de s^{(2)} + \frac{\de \s}{\sb'} \left( \de s' +\frac{\tb'}{2} \de \s \right) \right)' 
+ \tb' \left(\de \s^{(2)} - \frac{1}{2 \sb'} \de s \de s' \right)  \bigg] \\
&+ \h \eb_{\s I}  \ebs^M \ebs^N \ebsi^K \ebsi^L \Rb_{MKNL} \de s^2
+ \Gb_{LKI} \left[ \ebsi^L \left[ \ebsi^K \left(\de \s^{(2)} - \frac{1}{2 \sb'} \de s \de s' \right) \parenthnewlnln
+ \ebs^K \left( \de s^{(2)} + \frac{\de \s}{\sb'} \left( \de s' +\frac{\tb'}{2} \de \s \right) \right) \right]
+ \frac{1}{\sb'} \ebs^L \left(\ebsi^K \de \s + \ebs^K \de s \right) \left( \de s' + \tb' \de \s \right) \right]\\
&+ \ebsi^M \left(\ebsi^K \de \s + \ebs^K \de s \right) \left(\ebsi^L \de \s + \ebs^L \de s \right) \left[ \h \left(\bar{G}_{IP,M} - \bar{G}_{IM,P}\right) \Gb^P_{KL} \parenthnewln
- \Gb_{PIK} \Gb^P_{ML} + \frac{1}{4} \left( \bar{G}_{KL,IM} - \bar{G}_{IK,LM} - \bar{G}_{IL,KM} + 2 \bar{G}_{IM,KL} \right) \right],
\ea
\ee
In order to obtain the $s$-vielbeine we note that
\be
e_{s}^J = \de^J_{\; I} e_{s}^I = \left(\ebsi^J \bar{e}_{\s I} + \ebs^J \bar{e}_{s I} \right) e_{s}^I.
\ee
Expanding and rearranging the definitions
\be
G_{IJ} e_{\s}^I e_{\s}^J = G_{IJ} e_{s}^I e_{s}^J \equiv 1 \;\;\;\;\;\;\;\; \text{and} \;\;\;\;\;\;\;\; G_{IJ} e_{\s}^I e_{s}^J \equiv 0
\ee
up to second order, we obtain
\be \label{eq:deesJ}
\de e_{s}^J = \left(\ebsi^J \bar{e}_{\s I} + \ebs^J \bar{e}_{s I} \right) \de e_{s}^I = - \sb'^{-1} \left( \de s' + \tb' \de \s \right) \ebsi^J - \Gb^J_{KL} \ebs^L \left( \ebsi^K \de \s + \ebs^K \de s \right)
\ee 
at linear order.
Lowering the field space index gives
\be \label{eq:dees_I}
\ba
\de e_{s I} = - \sb'^{-1} \left( \de s' + \tb' \de \s \right) \eb_{\s I}  + \Gb_{LKI} \ebs^L \left( \ebsi^K \de \s + \ebs^K \de s \right).
\ea
\ee
At second order we have
\be\label{eq:de2esJ}
\ba
\de e_{s}^{(2) J} &= \left(\ebsi^J \bar{e}_{\s I} + \ebs^J \bar{e}_{s I} \right) \de e_{s}^{(2)I} \\
&=- \frac{1}{2 \sb'^2} \ebs^J \left( \de s' + \tb' \de \s \right)^2 - \frac{1}{\sb'} \ebsi^J\bigg[ - \sb'^{-1} \left( \de \s' - \tb' \de s \right) \left( \de s' + \tb' \de \s \right) \parenthnewlnbigg
+ \left( \de s^{(2)} + \frac{\de \s}{\sb'} \left( \de s' +\frac{\tb'}{2} \de \s \right) \right)' 
+ \tb' \left(\de \s^{(2)} - \frac{1}{2 \sb'} \de s \de s' \right)  \bigg] \\
&+  \ebs^M  \ebsi^K \left(  \h  \ebs^J \ebs^N \ebsi^L \de \s^2 + \ebsi^J \ebsi^N \ebs^L \de \s \de s \right) \Rb_{MKNL} 
- \Gb^J_{KL} \ebs^L \left[ \ebsi^K \left(\de \s^{(2)} - \frac{1}{2 \sb'} \de s \de s' \right) \parenthnewln
+ \ebs^K \left( \de s^{(2)} + \frac{\de \s}{\sb'} \left( \de s' +\frac{\tb'}{2} \de \s \right) \right) \right]
+ \sb'^{-1} \Gb^J_{KL} \left(\ebsi^L \de \s + \ebs^L \de s \right) \ebsi^K \left( \de s' + \tb' \de \s \right) \\
&+ \ebs^M \left(\ebsi^K \de \s + \ebs^K \de s \right) \left(\ebsi^L \de \s + \ebs^L \de s \right) \left[ - \h \p_M \Gb^J_{KL} + \Gb^J_{LN} \Gb^N_{MK} \right].
\ea
\ee
Lowering the field space index gives
\be\label{eq:de2es_I}
\ba
\de e_{s I}^{(2)}  &= 
- \frac{1}{2 \sb'^2} \eb_{s I} \left( \de s' + \tb' \de \s \right)^2 - \frac{1}{\sb'} \eb_{\s I} \bigg[ - \sb'^{-1} \left( \de \s' - \tb' \de s \right) \left( \de s' + \tb' \de \s \right) \parenthnewlnbigg
+ \left( \de s^{(2)} + \frac{\de \s}{\sb'} \left( \de s' +\frac{\tb'}{2} \de \s \right) \right)' 
+ \tb' \left(\de \s^{(2)} - \frac{1}{2 \sb'} \de s \de s' \right)  \bigg] \\
&+  \ebs^M  \ebsi^K \left(  \h  \eb_{s I} \ebs^N \ebsi^L \de \s^2 + \eb_{\s I} \ebsi^N \ebs^L \de \s \de s \right) \Rb_{MKNL} 
+ \Gb_{LIK} \ebs^L \left[ \ebsi^K \left(\de \s^{(2)} - \frac{1}{2 \sb'} \de s \de s' \right) \parenthnewln
+ \ebs^K \left( \de s^{(2)} + \frac{\de \s}{\sb'} \left( \de s' +\frac{\tb'}{2} \de \s \right) \right) \right]
- \sb'^{-1} \Gb_{LIK}  \left(\ebsi^K \de \s + \ebs^K \de s \right) \ebsi^L \left( \de s' + \tb' \de \s \right) \\
&+ \ebs^M \left(\ebsi^K \de \s + \ebs^K \de s \right) \left(\ebsi^L \de \s + \ebs^L \de s \right) \left[ \h \left(\bar{G}_{IP,M} - \bar{G}_{IM,P}\right) \Gb^P_{KL} \parenthnewln
- \Gb_{PIK} \Gb^P_{ML} + \frac{1}{4} \left( \bar{G}_{KL,IM} - \bar{G}_{IK,LM} - \bar{G}_{IL,KM} + 2 \bar{G}_{IM,KL} \right) \right].
\ea
\ee

\textbf{Lie derivative expansions:} 
Expanding the Lie derivative up to second order, we have for the fields
\be
\dot{\f}^I = u^0 \p_0 \f^I = \fb^{I'} + \de \f^{I'} - \fb^{I'} A^{(1)} + \de \f^{(2) I'} - \de \f^{I'} A^{(1)} - \fb^{I'} A^{(2)} + \frac{3}{2} \fb^{I'} A^{(1)2}.
\ee
Similarly, expanding $\dot{\s}^2$:
\be \label{eq:sdot2}
\ba
\dot{\s}^2 &\equiv G_{IJ} \dot{\f}^{I} \dot{\f}^{J} =
\sb'^2 
+2 \sb' \left( \de \s' - \tb' \de s - \sb' A^{(1)} \right) \\
&+ \sb'^2 \left( 4 A^{(1)2} - 2 A^{(2)} \right) - 4 \sb' A^{(1)} \left( \de \s' - \tb' \de s \right) + \left( \de \s' - \tb' \de s \right)^2  + \left( \de s' + \tb' \de \s \right)^2 \\
&+ 2 \sb' \left[ \de \s^{(2)} - \frac{1}{2 \sb'} \de s \de s' \right]' - 2 \sb' \tb'  \left[ \de s^{(2)} + \frac{\de \s}{\sb'} \left( \de s' + \frac{\tb'}{2} \de \s \right) \right] 
- \sb'^2  \ebs^I \ebs^J \ebsi^K \ebsi^L  \Rb_{IKJL} \de s^2  \\
&\stackrel{\de \s = 0}{\approx} \sb'^2 +2 \sb' \tb' \de s + 2 \sb' \tb' \de s^{(2)} - \Vb_{;ss} \de s^2 +\frac{\Vb_{,\s}}{\sb'} \de s  \de s', 
\ea
\ee
where the last expression is valid on large scales and in comoving gauge and where we have used the expressions for $A^{(1)}$ and $A^{(2)}$ given in (\ref{equ:A1}) and (\ref{equ:A2}), respectively.

We can then compute the perturbation expansion in $\dot{\s} = \dot{\s}^{(0)}+ \de \dot{\s}^{(1)} + \de \dot{\s}^{(2)} + \dots$:
\be
\dot{\s}^{(0)} = \sqrt{ \left(\dot{\s}^2\right)^{(0)} } = \sqrt{\bar{G}_{IJ} \fb^{I'} \fb^{J'} } \equiv \sb'
\ee
at zeroth order,
\be
\ba
\de \dot{\s}^{(1)} &= \frac{\de \! \left(\dot{\s}^2\right)^{(1)}}{2 \dot{\s}^{(0)}} 
= \de \s' - \tb' \de s - \sb' A^{(1)} \stackrel{\de \s = 0}{\approx} \tb' \de s
\ea
\ee
at linear order, and
\be
\ba
 \de \dot{\s}^{(2)} &= \frac{\de \! \left(\dot{\s}^2\right)^{(2)} - \left( \de \dot{\s}^{(1)} \right)^2}{2 \dot{\s}^{(0)}}  \\
&= \left[ \de \s^{(2)} - \frac{1}{2 \sb'} \de s \de s' \right]' - \tb' \left[ \de s^{(2)} + \frac{\de \s}{\sb'} \left( \de s' + \frac{\tb'}{2} \de \s \right) \right]  + \h \sb'^{-1} \left( \de s' + \tb' \de \s \right)^2 \\
&-\left( \de \s' - \tb' \de s \right)  A^{(1)} - \sb' A^{(2)} + \frac{3}{2} \sb' A^{(1)2} - \h \sb'^2  \ebs^I \ebs^J \ebsi^K \ebsi^L  \Rb_{IKJL} \de s^2 \\
&\stackrel{\de \s = 0}{\approx} \tb' \de s^{(2)} + \frac{\Vb_{,\s}}{2 \sb'^2} \de s \de s' - \frac{1}{2 \sb'} \left( \Vb_{;ss}  + \tb'^2 \right) \de s^2
\ea
\ee
at quadratic order.

\textbf{Metric perturbations $A^{(1)}$ and $A^{(2)}$:} \label{app:A1A2}
To determine $A^{(1)}$ and $A^{(2)}$ we make use of the fact that on large scales the comoving energy density perturbation is zero, $\de \ep \approx 0$. Moreover, in comoving gauge, it simplifies to $\de \r$:
\be
\de \ep \equiv \de \r - \frac{\rb'}{\sb'} \de \s \stackrel{\de \s =0}{=} \de \r \approx 0
\ee
at first order, and
\be
\de \ep^{(2)} \equiv \de \r^{(2)} - \frac{\rb'}{\sb'} \de \s^{(2)} - \frac{\de \s}{\sb'} \left[ \de \ep' + \h \left(\frac{\rb'}{\sb'} \right)' \de \s + \frac{\rb'}{\sb'} \tb' \de s \right] \stackrel{\de \s =0}{=} \de \r^{(2)} \approx 0
\ee
at second order.

The comoving energy density is given by (\ref{eq:rho}), and can be expanded up to second order:
\be
\ba
\r = \, &\h G_{IJ} \dot{\f}^I \dot{\f}^J + \h G_{IJ} g^{ij} \na_i \f^I \na_j \f^J + V 
\approx \h \sd^2 + V \\
\therefore \; \rb \approx \, &\h \sb'^2 + \Vb  \\
\therefore \; \de \r \approx \, & -2 \sb'  \tb' \de s  - \sb'^2 A^{(1)}  \\
\therefore \; \de \r^{(2)} \approx \, &-2 \sb'  \tb' \de s^{(2)}   - \sb'^2 A^{(2)}   - \frac{\Vb_{,\s}}{\sb'} \de s \de s' + \left( \Vb_{;ss} + 2 \tb'^2 \right) \de s^2  + 2 \sb' A^{(1)} \left( \tb' \de s + \sb' A^{(1)} \right),
\ea
\ee
where we have neglected spatial gradients.

At linear order, we thus have
\be \label{equ:A1}
A^{(1)} \approx -2 \frac{\tb'}{\sb'} \de s,
\ee
and at second order
\be \label{equ:A2}
\ba
A^{(2)} \approx &-2  \frac{\tb'}{\sb'} \de s^{(2)}  +  \frac{1}{\sb'^2} \left( \Vb_{;ss} + 2 \tb'^2 \right) \de s^2  - \frac{\Vb_{,\s}}{\sb'^3} \de s \de s' + 2 A^{(1)} \left( \frac{\tb'}{\sb'} \de s + A^{(1)} \right) \\
\approx &-2  \frac{\tb'}{\sb'} \de s^{(2)}  + \frac{1}{\sb'^2} \left( \Vb_{;ss} + 6 \tb'^2 \right) \de s^2  - \frac{\Vb_{,\s}}{\sb'^3} \de s \de s'.
\ea
\ee
\\

\textbf{Perturbations of other important quantities:} 
\be \label{eq:deV1}
\ba
\de V^{(1)} &= \Vb_{,\s}  \de \s  - \sb' \tb' \de s  \stackrel{\de \s = 0}{\approx}  - \sb' \tb' \de s
\ea
\ee

\be \label{eq:deV2}
\ba
\de V^{(2)} &= \Vb_{,\s} \left[ \de \s^{(2)} - \frac{1}{2 \sb'} \de s \de s' \right] - \sb' \tb' \left[ \de s^{(2)} + \frac{\de \s}{\sb'} \left( \de s' +\frac{\tb'}{2} \de \s \right) \right] \\
&+ \h \Vb_{;\s\s} \de \s^2 + \Vb_{;s\s} \de \s \de s + \h \Vb_{;ss} \de s^2 \\
&\mkern-16mu \stackrel{\de \s = 0}{\approx}  - \sb' \tb' \de s^{(2)} -  \frac{\Vb_{,\s} }{2 \sb'} \de s \de s' + \h \Vb_{;ss} \de s^2
\ea
\ee

\be \label{eq:deV3}
\ba
\de V^{(3)} &\stackrel{\de \s = 0}{\approx}  - \sb' \tb' \left[\de s^{(3)} - \frac{1}{6 \sb'^2} \de s \de s'^2 \right] - \Vb_{,\s} \left[ \frac{1 }{2 \sb'} \left(\de s \de s^{(2)} \right)' + \frac{\tb'}{6 \sb'^2} \de s^2 \de s' \right] \\
&\;\;\;\, +\frac{1}{6} \Vb_{;sss} \de s^3 + \Vb_{;ss} \de s \de s^{(2)} - \frac{1}{2 \sb'} \Vb_{;s \s} \de s^2 \de s'  
\ea
\ee

\be
\de V_{;ss} = \Vb_{;sss} \de s - 2 \frac{\Vb_{;s\s}}{\sb'} \de s',
\ee
with $\Vb_{;sss} = \ebs^I \ebs^J \ebs^K \Vb_{;IJK} $.
\be
\ba
\de V_{;ss}^{(2)} &\approx \Vb_{;sss} \de s^{(2)} + \h \Vb_{;ssss} \de s^2 - \frac{5}{2 \sb'} \Vb_{;ss \s} \de s \de s' + \frac{\Vb_{;\s\s} - \Vb_{;ss}}{\sb'^2} \de s'^2 -  \frac{2}{\sb'} \Vb_{;s\s} \left( \de s^{(2)'} + \frac{\tb'}{2 \sb'} \de s \de s' \right) \\
&-\frac{2 \Vb_{,\s}}{\sb'} \ebs^I \ebs^J \ebsi^K \ebsi^L \Rb_{IKJL} \de s \de s',
\ea
\ee
with $\Vb_{;ssss} = \ebs^I \ebs^J \ebs^K \ebs^L \Vb_{;IJKL} $.

\be
\ba
\de \dot{\th} \approx - \frac{\Vb_{;ss}}{\sb'} \de s + \frac{\Vb_{,\s}}{\sb'^2} \de s' - \frac{\tb'^2}{\sb'} \de s.
\ea
\ee
\be
\ba
\de \dot{\th}^{(2)} &\approx  \frac{\Vb_{,\s}}{\sb'^2} \de s^{(2)'} - \frac{1}{\sb'} \left( \Vb_{;ss} + \tb'^2 \right) \de s^{(2)} - \frac{\tb'}{2 \sb'^2} \de s'^2 + \frac{1}{2 \sb'^2} \left( 4 \frac{\tb' \Vb_{,\s}}{\sb'} - 3 \tb'' + 9 H \tb' \right) \de s \de s'\\
&+ \frac{1}{2 \sb'^2}\left( - \sb' \Vb_{;sss} + 3 \Vb_{;ss} \tb' + \tb'^3  \right) \de s^2.
\ea
\ee

\be
\de \left[\dot{\s}^2e_s^I e_s^J e_{\s}^K e_{\s}^L R_{IKJL}  \right] \approx \ebs^I \ebs^J \ebsi^K \ebsi^L  \left[2 \sb' \tb' \bar{R}_{IKJL} + \sb'^2 \ebs^N \cD_N \bar{R}_{IKJL}  \right] \de s
\ee
\be
\ba
\de \left[\dot{\s}^2e_s^I e_s^J e_{\s}^K e_{\s}^L R_{IKJL}  \right]^{(2)} &\approx \ebs^I \ebs^J \ebsi^K \ebsi^L  \left[\bar{R}_{IKJL} \left(2 \sb' \tb' \de s^{(2)}  + \frac{\Vb_{,\s}}{\sb'} \de s \de s' - \Vb_{;ss} \de s^2 \right) \parenthnewln
+ \cD_N \bar{R}_{IKJL}  \left( \sb'^2 \ebs^N  \de s^{(2)} - \frac{\sb'}{2 } \ebsi^N \de s \de s'  + 2 \sb' \tb' \ebs^N  \de s^2 \right) \parenthnewln
+ \sb'^2 \ebs^N \ebs^Q \left(   \h \cD_Q \cD_N \bar{R}_{IKJL} -  \Rb_{IKJP} \Rb^P_{\; NLQ} +  \Rb_{IKJL} \Rb^P_{\; NPQ} \right) \de s^2\right] 
\ea
\ee

\textbf{Useful derivatives:} 
\be
\tb'' = - \Vb_{;s\s} + 3 H \tb' + 2 \frac{\tb' \Vb_{,\s}}{\sb'}
\ee
\be
\Vb_{,\s}' = \sb' \left(\Vb_{;\s\s} - \tb'^2 \right)
\ee 
\be
\Vb_{;ss}' = \sb' \Vb_{;ss\s} - 2 \tb' \Vb_{;s\s}
\ee
\be
\ebs^{J'} = -  \tb' \ebsi^J - \Gb^J_{KL} \sb' \ebs^K \ebsi^L
\ee
\be
\left[\ebs^I \ebs^J \ebsi^K \ebsi^L \Rb_{IKJL} \right]' = \ebs^I \ebs^J \ebsi^K \ebsi^L  \sb' \ebsi^M \cD_M \Rb_{IKJL}
\ee

\section{Simplifications for our specific model} \label{sec:specificmodel}
In our specific model, the metric and its inverse are given by
\be
G_{IJ} = \begin{pmatrix}
  1  &  0 \\
  0  &  \O(\f)^2  \\
 \end{pmatrix},
\;\;\;\;\; \text{and} \;\;\;\;\;
G^{IJ} = \begin{pmatrix}
  1  &  0 \\
  0  &  \O(\f)^{-2}  \\
 \end{pmatrix}.
\ee
The non-trivial connections derived from this metric are then
\be
\ba
\Gb^{\phi}_{\chi \chi} =  &-\O \O_{,\f} 
\ea
\ee
and
\be
\ba
\Gb^{\chi}_{\phi \chi} =&\textcolor{white}{+} \O^{-1} \O_{,\f},
\ea
\ee
while the only non-trivial component (up to those related by symmetry) of the Riemann tensor is
\be
 \Rb_{\phi \chi \phi \chi} = - \O \O_{, \phi \phi}.
\ee
The covariant derivatives of the Riemann tensor in our model are
\be
\cD_{\chi}  \Rb_{\phi \chi \phi \chi} = 0,
\ee
\be
\cD_{\f} \Rb_{\phi \chi \phi \chi} = \O_{,\f} \O_{, \phi \phi} - \O \O_{, \f\f\f},
\ee
\be
\cD_{\f} \cD_{\f} \Rb_{\f \chi \f \chi} = - \O \O_{, \f\f\f\f} +2 \O_{,\f} \O_{, \f \f\f} +\O_{, \f\f}^2 - 2 \O^{-1} \O_{,\f}^2 \O_{, \f\f},
\ee
\be
\cD_{\chi} \cD_{\chi} \Rb_{\phi \chi \phi \chi} = - \O^2 \O_{,\f} \O_{, \f \f\f} + \O \O_{,\f}^2 \O_{, \f \f},
\ee
\be
\cD_{\f} \cD_{\chi} \Rb_{\phi \chi \phi \chi} = \cD_{\chi} \cD_{\f} \Rb_{\phi \chi \phi \chi} = 0.
\ee
We can define the zweibeine, via $e^I_{\s} \equiv \frac{\dot{\f}^J}{\dot{\s}}$, such that
\be \label{eq:vielbeinsigma1}
\eb_{\s}^{\f} = \frac{\fb'}{\sb'}, \;\;\;\;\; \eb^{\chi}_{\s} = \frac{\bar{\chi}'}{\sb'},
\ee
\be \label{eq:ephis}
\eb^{\f}_{s} = - \O \frac{\bar{\chi}'}{\sb'}, \;\;\;\;\; \eb^{\chi}_{s} = \O^{-1} \frac{\fb'}{\sb'},
\ee
\be
\eb_{\s \f} = \frac{\fb'}{\sb'}, \;\;\;\;\; \eb_{\s \chi} = \O^2 \frac{\bar{\chi}'}{\sb'},
\ee
\be \label{eq:vielbeins2}
\eb_{s \f} = - \O \frac{\bar{\chi}'}{\sb'}, \;\;\;\;\; \eb_{s \chi} = \O \frac{\fb'}{\sb'}.
\ee
Note that within this setup we must take $\eb_{\s}^{\f}=-1$ during the ekpyrotic phase; this is because $\s$ is defined to increase along the background trajectory \cite{RenauxPetel:2008gi} and thus $\sb'=-\fb'$ is the velocity of the background trajectory in the constant $\chi$ backgrounds that we are interested in.

\textbf{Simplifications during the ekpyrotic phase:} 

\be
\de s |_{\text{ekp}}  = - \O \de \chi
\ee
\be
\de s' |_{\text{ekp}}  = - \fb' \O_{,\f} \de \chi = - \sb' \O^{-1} \O_{,\f} \de s \;\;\;\;\; \because  \;\;\; \de \chi' |_{\text{ekp}}  = 0
\ee
\be
\de s^{(2)} |_{\text{ekp}}  = 0
\ee

\be
\Vb_{,\s} |_{\text{ekp}}  = - \Vb_{,\f}
\ee
\be
\Vb_{,s} |_{\text{ekp}}  = \tb' = 0
\ee
\be
\Vb_{;\s\s} |_{\text{ekp}}  = \Vb_{,\f\f}
\ee
\be
\Vb_{;s\s} |_{\text{ekp}}  = 0
\ee
\be
\Vb_{;ss} |_{\text{ekp}}  = \O^{-1} \O_{,\f} \Vb_{,\f}
\ee
\be
\Vb_{;ss\s} |_{\text{ekp}}  = \left( - \O^{-1} \O_{,\f\f}  + \O^{-2} \O_{,\f}^2 \right) \Vb_{,\f} - \O^{-1} \O_{,\f} \Vb_{,\f\f}
\ee
\be
\Vb_{;sss} |_{\text{ekp}}  = 0
\ee
\be
\Vb_{;ssss} |_{\text{ekp}}  = -3  \O^{-3} \O_{,\f}^3 \Vb_{,\f}  + 3  \O^{-2} \O_{,\f}^2 \Vb_{,\f\f} +  \O^{-2} \O_{,\f} \O_{,\f\f} \Vb_{,\f} 
\ee

\be
\cD_{\chi} \Rb_{ \chi \f \chi \f} |_{\text{ekp}}  = 0
\ee
\be
\cD_{\f} \Rb_{ \chi \f \chi \f} |_{\text{ekp}}  = \O_{,\f} \O_{,\f\f} -  \O  \O_{,\f\f\f}
\ee
\be
\cD_{\chi} \cD_{\chi} \Rb_{ \chi \f \chi \f} |_{\text{ekp}}  = \O \O_{,\f} \left( \O_{,\f} \O_{,\f\f} -  \O  \O_{,\f\f\f} \right)
\ee

\be
A^{(1)} |_{\text{ekp}}=  A^{(2)} |_{\text{ekp}} = 0,
\ee
\be
\de \sd^2 |_{\text{ekp}}= \de \left(\sd^2 \right)^{(2)} |_{\text{ekp}} = 0,
\ee
\be
\bar{\Theta} = 3H, \;\;\;\; \de \Theta |_{\text{ekp}}= \de \Theta^{(2)} |_{\text{ekp}} = 0,
\ee
\be
\de V_{;ss} |_{\text{ekp}}= \de V_{;ss}^{(2)} |_{\text{ekp}} = 0,
\ee
\be
\de \td |_{\text{ekp}}= \de \td^{(2)} |_{\text{ekp}} = 0,
\ee
\be
\de \left[ \sd^2 e_s^I e_s^J e_{\s}^K e_{\s}^L R_{IKJL} \right] |_{\text{ekp}}= \de \left[ \sd^2 e_s^I e_s^J e_{\s}^K e_{\s}^L R_{IKJL} \right]^{(2)} |_{\text{ekp}} = 0,
\ee
\be
\de s_i^{(2)} |_{\text{ekp}} = \p_i \de s^{(2)} |_{\text{ekp}} =  0,
\ee
\be
\de s_i^{(3)} |_{\text{ekp}} = \dot{\de s_i}^{(3)} |_{\text{ekp}}= \ddot{\de s_i}^{(3)} |_{\text{ekp}} = 0.
\ee

\be
\de e_{\s}^{\f}|_{\text{ekp}}= \de e_{\s \f}|_{\text{ekp}}= 0,
\ee
\be
\de e_{\s}^{\chi}|_{\text{ekp}}= \de e_{\s \chi}|_{\text{ekp}}= 0,
\ee
\be
\de e_{s}^{\f}|_{\text{ekp}}= \de e_{s \f}|_{\text{ekp}}= 0,
\ee
\be
\de e_{s}^{\chi}|_{\text{ekp}}= \de e_{s \chi}|_{\text{ekp}}= 0,
\ee
\be
\de e_{\s}^{\f (2)}|_{\text{ekp}}= \de e_{\s \f}^{\, (2)}|_{\text{ekp}}= 0,
\ee
\be
\de e_{\s}^{\chi (2)}|_{\text{ekp}}= \frac{\tb'}{2 \sb'} \O^{-2} \O_{,\f} \de s^2, \;\;\;\;\;\;
\de e_{\s \chi }^{\, (2)}|_{\text{ekp}}= \frac{\tb'}{2 \sb'} \O_{,\f} \de s^2,
\ee
\be
\de e_{s}^{\f (2)}|_{\text{ekp}}= \de e_{s \f }^{\, (2)}|_{\text{ekp}}= - \frac{\tb'}{2 \sb'} \O^{-1} \O_{,\f} \de s^2,
\ee
\be
\de e_{s}^{\chi (2)}|_{\text{ekp}}= \de e_{s \chi}^{\, (2)}|_{\text{ekp}}= 0.
\ee

\be
\de G_{IJ}|_{\text{ekp}} = 0,
\ee
\be
\de G_{IJ}^{\,(2)}|_{\text{ekp}} = 0.
\ee

\bibliographystyle{apsrev}
\bibliography{NonMinimalEntropicBib.bib}

\end{document}